\documentclass[twocolumn,journal]{IEEEtran}
\usepackage[T1]{fontenc}
\usepackage[latin9]{inputenc}
\usepackage{amsbsy}
\usepackage{graphicx}
\usepackage{esint}
\usepackage[unicode=true,
bookmarks=true,bookmarksnumbered=true,bookmarksopen=true,bookmarksopenlevel=1,
breaklinks=false,pdfborder={0 0 1},backref=false,colorlinks=false]
{hyperref}
\hypersetup{pdftitle={Parametric Modeling of Serpentine Waveguide Traveling Wave Tubes},
	pdfauthor={Kasra Rouhi},
	pdfpagelayout=OneColumn, pdfnewwindow=true, pdfstartview=XYZ, plainpages=false}

\usepackage{fancyhdr}

\makeatletter

\providecommand{\tabularnewline}{\\}

\let\oldforeign@language\foreign@language
\DeclareRobustCommand{\foreign@language}[1]{%
	\lowercase{\oldforeign@language{#1}}}

\usepackage[caption=false,font=footnotesize]{subfig}

\makeatother

\begin{document}
	\title{Parametric Modeling of Serpentine Waveguide Traveling Wave Tubes}
	\author{Kasra~Rouhi, Robert Marosi, Tarek Mealy, Alexander Figotin, and Filippo Capolino, \thanks{Kasra Rouhi, Robert Marosi, Tarek Mealy, and Filippo Capolino are
			with the Department of Electrical Engineering and Computer Science,
			University of California, Irvine, CA 92697 USA, e-mails: \protect\href{mailto:kasra.rouhi@uci.edu}{kasra.rouhi@uci.edu},
			\protect\href{mailto:rmarosi@uci.edu}{rmarosi@uci.edu}, \protect\href{mailto:tmealy@uci.edu}{tmealy@uci.edu}
			and \protect\href{mailto:f.capolino@uci.edu}{f.capolino@uci.edu}.}\thanks{Alexander Figotin is with the Department of Mathematics, University
			of California, Irvine, CA 92697 USA, e-mail: \protect\href{mailto:afigotin@uci.edu}{afigotin@uci.edu}.}}
	
	\maketitle

	\thispagestyle{fancy}

\begin{abstract}
A simple and fast model for numerically calculating small-signal gain
in serpentine waveguide traveling-wave tubes (TWTs) is described.
In the framework of the Pierce model, we consider one-dimensional
electron flow along a dispersive single-mode slow-wave structure (SWS),
accounting for the space-charge effect. The analytical model accounts
for the frequency-dependent phase velocity and characteristic impedance
obtained using various equivalent circuit models from the literature,
validated by comparison with full-wave eigenmode simulation. The model
includes a relation between the modal characteristic impedance and
the interaction (Pierce) impedance of the SWS, including also an extra
correction factor that accounts for the variation of the electric
field distribution and hence of the interaction impedance over the
beam cross section. By applying boundary conditions to our generalized
Pierce model, we compute both the theoretical gain of a TWT and all
the complex-valued wavenumbers of the hot modes versus frequency and
compare our results with computationally intensive particle-in-cell
(PIC) simulations; the good agreement in the comparison demonstrates
the accuracy and simplicity of our generalized model. For various
examples where we vary the average electron beam (e-beam) phase velocity,
average e-beam current, number of unit cells, and input radio frequency
(RF) power, we demonstrate that our model is robust in the small-signal
regime. The purpose of this paper is not to design a TWT with performance
that competes with previous ones, but to develop an accurate and simple
model to predict TWT performance that can be used as a design tool.
\end{abstract}


\IEEEpeerreviewmaketitle{}

\section{Introduction}

\IEEEPARstart{T}{he} traveling-wave tube (TWT) is a type of common
microwave vacuum electron tube that has been widely used for applications
such as communication, radar, and electronic countermeasures \cite{han2005investigations,wong2020recent,benford2015high,paoloni2021millimeter}.
Among the different kinds of TWTs, the serpentine waveguide TWT has
advantages over other kinds of millimeter wave TWTs (e.g. helix TWT,
coupled cavity (CC) TWT, ring-bar TWT) due to its moderate bandwidth
with power-handling capacity at higher frequencies and its compatibility
with planar fabrication using lithography or micromachining \cite{dohler1987serpentine,liu1995folded,collins1998technique,collins1999new,gong2011experimental}.
The slow-wave structure (SWS) of the serpentine waveguide TWT is formed
by bending rectangular waveguides in the electric field plane ($E$-plane).
Also, a cylindrical electron beam (e-beam) is transported through
the cylindrical beam tunnel to interact with the radio frequency (RF)
propagating wave. Although the serpentine waveguide SWS's performance
is limited by its low interaction impedance and interaction efficiency,
many schemes of enhancing the on-axis interaction impedance and also
enhancement of interaction efficiency have been proposed \cite{he2010investigation,liao2010rectangular,tian2011novel,hou2013novel,wei2014novel,marosi2022Three}.

In order to analyze the e-beam and electromagnetic (EM) wave dynamics
of a serpentine waveguide TWT, it is necessary to examine the EM characteristics
of the SWS. Various analytical models have been developed for its
characterization. In 1987, Dohler et al. proposed a simple analytical
method for determining the dispersion characteristics and the interaction
impedance of the EM modes in the serpentine waveguide \cite{dohler1987serpentine}.
Liu suggested an analytical formulation adding the effect of bends
\cite{liu1995folded}. Then, researchers developed a closed-form algebraic
dispersion relation based on an equivalent circuit model that also
considered the effect of mismatch between straight and bend sections
as well as an approximate model for beam holes \cite{choi1995folded,ha1998theoretical}.
A thorough equivalent circuit analysis of serpentine waveguides by
modeling the effect of beam tunnels as orthogonal stubs was developed
by Booske et al. \cite{booske2005accurate} for the calculation of
dispersion characteristics, following the approach of transmission
line (TL) cascading networks and benchmarked using three-dimensional
(3D) simulations with Ansys HFSS, MAFIA, and CST Studio Suite. Recently,
Antonsen et al. \cite{antonsen2013transmission} developed a hybrid
model consisting of a combination of TL segments and lumped electrical
elements, which is utilized to analyze serpentine waveguide dispersion
characteristics and interaction impedance. The model also captures
the effects of asymmetric fields and beam tunnel misalignment. Although
some commercial full-wave simulation software like Ansys HFSS and
CST Studio Suite are versatile and can analyze SWS characteristics,
simulation times are longer than analytical methods. Therefore, analytical
methods are preferred for quickly iterating through and optimizing
various SWS designs.

\begin{figure*}[tbh]
\begin{centering}
\includegraphics[width=0.93\textwidth]{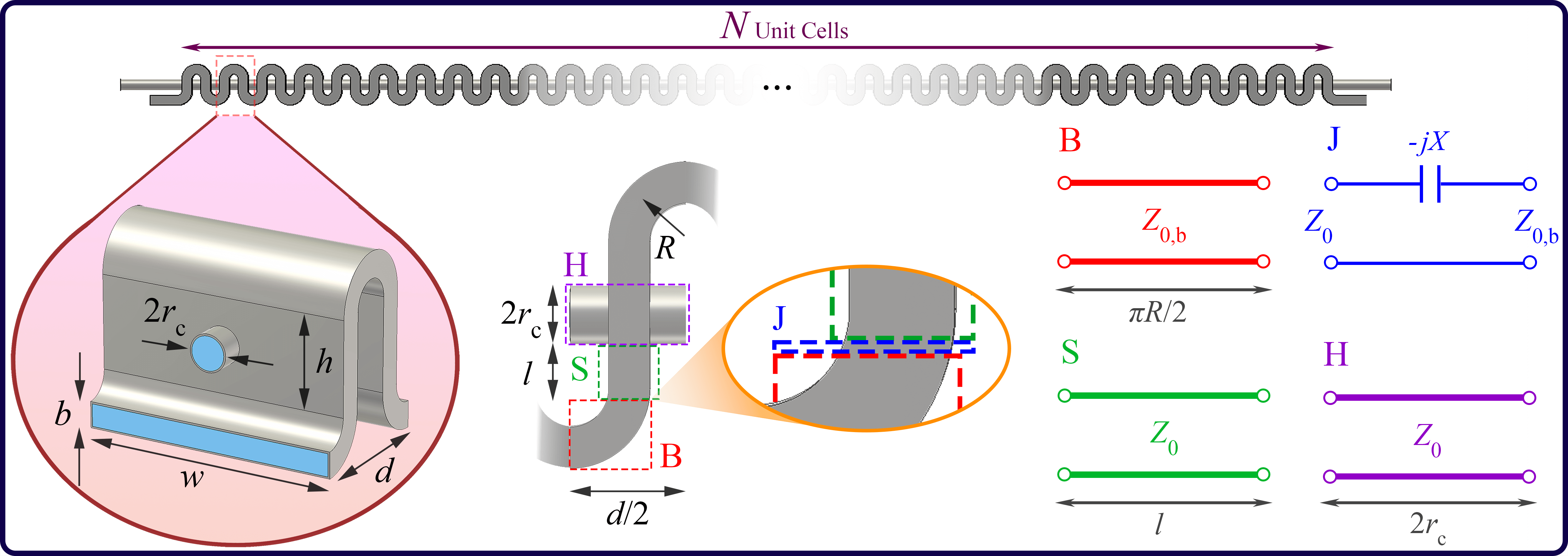}
\par\end{centering}
\centering{}\caption{Schematic illustration of a serpentine waveguide unit cell, constitutive
segments (colored dashed rectangles), and parametric dimensions are
shown in the left and central panels. The equivalent TL model for
the TE10 mode in each segment is shown in the right panel for: (B)
$E$-plane circular bend, (J) circular bend to straight waveguide
junction, (S) straight waveguide section, and (H) e-beam hole.\label{fig: Main}}
\end{figure*}

To design and analyze serpentine waveguide TWTs, various beam-EM wave
interaction models exist. Particle-in-cell (PIC) simulations are widely
used to characterize the beam-EM wave interaction of TWTs because
they predict amplification performance. Nevertheless, the computational
burden of 3D PIC simulators is high compared to other TWT codes. The
United States Naval Research Laboratory applied the hybrid TL model
to the large signal beam-EM wave interaction programs (CHRISTINE-CC
and TESLA-CC), which are used for analyzing CC-TWTs. Then, they extended
a 1D frequency-domain interaction model named CHRISTINE-FW, developed
for folded waveguide TWTs \cite{chernin20141} and a two-dimensional
(2D) frequency-domain interaction model named TESLA-FW \cite{chernyavskiy2013parallel,chernyavskiy2016large}.
Also, a large signal beam-EM wave interaction code with computational
efficiency improvements was developed by Meyne et al. \cite{meyne2016large}.
A 3D steady-state beam-EM wave interaction code using a three-port
network representation of the circuit and a set of discrete ray representations
of the 3D e-beam was developed by Yan et al. \cite{yan20163}. In
addition to previous models, a nonlinear model for the numerical simulation
of terahertz serpentine waveguide TWT is described in \cite{li2015nonlinear},
in which the propagated EM wave in the SWS is represented as an infinite
set of space harmonics that interact with an e-beam. Also, an improved
large-signal model was developed in \cite{zhang2020active}, which
predicts beam-EM wave interaction with an analytical method. Recently,
Figotin \cite{figotin2023analytic} advanced a Lagrangian field theory
of CC-TWTs that integrates into it the space-charge effects; that
model can also be used for serpentine waveguide TWTs as explained
in detail in that paper.

In this paper, we present an analytical model for analyzing beam-EM
wave interactions in serpentine waveguide TWTs shown in Fig. \ref{fig: Main}.
We develop a model that can be used to obtain the small-signal gain
and the \textquotedbl hot eigenmodes\textquotedbl{} dispersion, accounting
for nonuniform beam-EM wave interaction. We refer to the modes of
the interactive system, where the e-beam interacts with the EM wave
of the SWS, as \textquotedblleft hot modes\textquotedbl{} or ``hot
eigenmodes'', which are complex modes, with each hot eigenmode composed
of both EM and space-charge waves. First, we show various methods
from the literature that can be used to calculate SWS cold characteristics,
i.e., characteristic impedance and phase velocity, based on the equivalent
circuit model presented in \cite{booske2005accurate}. We calculate
the interaction impedance, which is one of the critical parameters
for predicting TWT gain. Based on the fundamental equations of the
Pierce model \cite{pierce1947theory,pierce1949new,pierce1950Book,pierce1951waves},
we further develop the model to account for frequency-dependent parameters
and the space-charge effect, following the method explained in \cite{rouhi2021exceptional}
for a helix TWT. Then, we introduce the frequency-dependent coupling
strength coefficient which shows the strength of the interaction between
e-beam and EM wave and also connects interaction impedance and characteristic
impedance. We also include the small frequency-independent factor
$\delta_{\mathrm{e}}$ that corrects for the nonuniform interaction
impedance over the beam cross section. This correction factor models
the nonuniform interaction between the EM wave and the e-beam in the
interaction gap. Moreover, we model the e-beam effect on the equivalent
TL model by using the electronic beam admittance per unit length $Y_{\mathrm{b}}$,
accounting for the space-charge effect. By introducing $Y_{\mathrm{b}}$,
it is possible to find out the conditions that lead to amplification
in the TWT system. Finally, we utilize the proposed theoretical method
to predict the gain versus frequency of a TWT amplifier and we compare
our results to those from computationally intensive 3D PIC simulations,
showing high accuracy. In order to show the flexibility and accuracy
of our method, comparison with 3D PIC simulations for many examples
is done by varying the e-beam parameters such as the average e-beam
phase velocity, average e-beam current, number of unit cells, and
input RF power.

The organization of this paper is as follows. In Section \ref{sec:Statement-of-Main},
we highlight the main achievements of our developed model. Then, we
show how to combine some analytical methods from the literature to
calculate the cold parameters of the serpentine waveguide in Section
\ref{sec:Equivalent-Circuit-Model}. An example of a cold model characteristic
calculation is presented in Section \ref{sec:Validation-of-Equivalent}.
We describe the conventional method to calculate interaction impedance
and introduce the extra correction factor $\delta_{\mathrm{e}}$ required
for our model in Section \ref{sec:Pierce-Impedance}. We develop a
model for beam-EM wave interaction in Section \ref{sec:e-beam-and-EM}
and evaluate it by providing an example in Section \ref{sec:Validation-of-Hot},
where we apply boundary conditions to determine the TWT gain. Next,
we demonstrate the accuracy and efficiency of our model in Section
\ref{sec:Parameter-Study} by varying TWT parameters. Finally, we
conclude the paper in Section \ref{sec:Conclusion} and discuss the
supplementary information in the Appendices.

\section{Summary of Main Results\label{sec:Statement-of-Main}}

\begin{figure*}[tbh]
\begin{centering}
\includegraphics[width=1\textwidth]{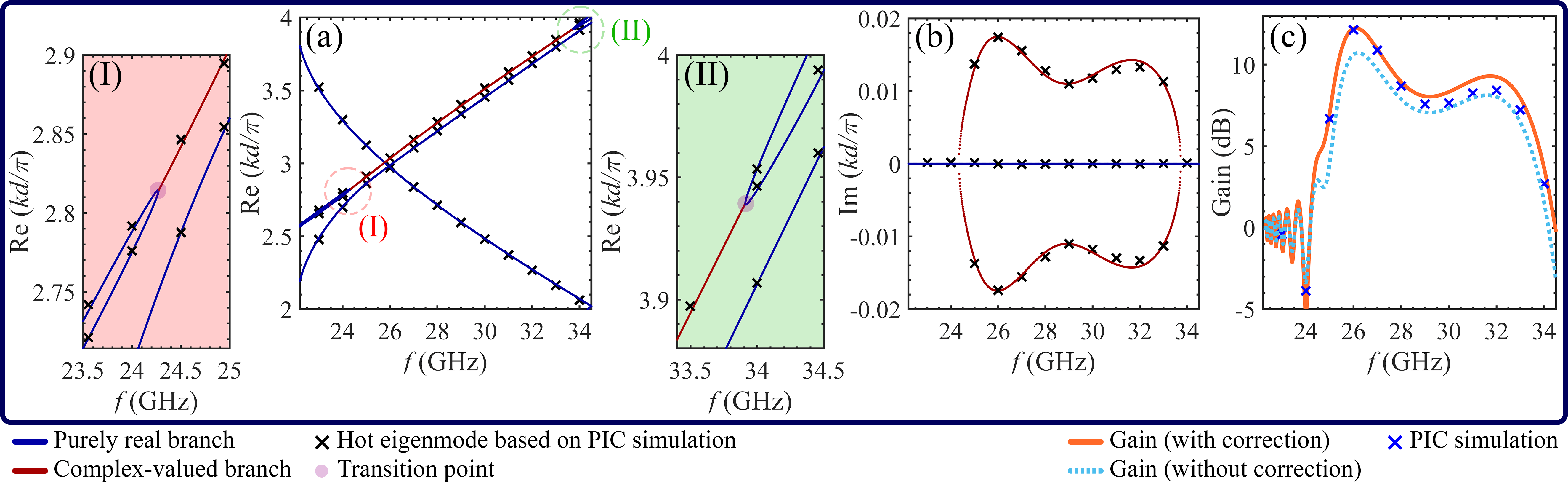}
\par\end{centering}
\centering{}\caption{The (a) real and (b) imaginary parts of complex-valued wavenumbers
of the hot modes, where dark blue curves indicate branches with purely
real wavenumbers, dark red curves indicate branches with complex-valued
wavenumbers, and black crosses indicate the results obtained using
a hot eigenmode solver for beam-loaded SWS based on PIC simulations
\cite{mealy2022traveling}. (I) and (II) show the real part of the
complex-valued wavenumbers of the hot modes near the two transition
points (light purple circles). (c) The theoretical gain (with/without
correction factor $\delta_{\mathrm{e}}$) is compared with that from
PIC simulation. The parameters used for this example are provided
in Sections \ref{sec:Validation-of-Equivalent} and \ref{sec:Validation-of-Hot}.\label{fig: MainResults}}
\end{figure*}
We present a summary of the main results calculated by our developed
model and compared to PIC simulations, leaving explanations, technical
details and numerical examples in the sections that follow. In our
developed model, we introduce an additional correction factor $\delta_{\mathrm{e}}$
that accounts for transverse variations in the axial electric field
distribution that affect the average interaction impedance over the
beam cross section. As a result of this correction factor, we can
model the nonuniform interactions between the EM wave and the e-beam
in the interaction gap. Figures \ref{fig: MainResults}(a) and (b)
illustrate the real and imaginary parts of complex-valued wavenumbers
of the eigenmodes supported by the serpentine waveguide with the e-beam,
i.e., of the hot modes, for the example with the parameters provided
in Sections \ref{sec:Validation-of-Equivalent} and \ref{sec:Validation-of-Hot}.
The solid lines in Figs. \ref{fig: MainResults}(a) and (b) represent
the calculated frequency dispersion of the hot modes resulting from
the interaction between the guided EM wave in the SWS and the two
space charge waves of the e-beam. The dark blue curves indicate ``stable
branches'' whose imaginary parts of the wavenumber of the hot modes
are equal to zero and hence are not amplified. In contrast, dark red
curves indicate branches whose imaginary parts of the wavenumber are
nonzero, and the positive values of the imaginary part allow for amplification
(unstable or amplification branch). In order to verify our theoretical
calculations displayed by solid curves, we calculate the real and
imaginary parts of the complex-valued wavenumbers of the hot modes
at a discrete set of frequencies by using the \textquotedblleft hot
eigenmode solver\textquotedblright{} for beam-loaded SWS based on
PIC simulations developed in \cite{mealy2022traveling} (indicated
by black crosses). This eigenmode solver is based on accurate PIC
simulations of finite-length hot structures, which consider the precise
SWS geometry, the EM properties of the materials, the cross-sectional
area of the e-beam, the confining magnetic field, and the space-charge
effect. The advantage of the hot eigenmode solver is that the use
of PIC simulations allows us to find the hot eigenmodes that fully
account for all physical aspects of the problem without the need to
rely on intermediate parameters, such as the interaction impedance
or plasma frequency reduction factor used in other solvers \cite{mealy2022traveling}.
There is excellent agreement between our theoretical model and the
PIC-based eigenmode solver of \cite{mealy2022traveling}, both in
the real and imaginary parts of the complex wavenumber. In addition,
we show the zoomed-in plot of the real part of the complex-valued
wavenumber near the two transition points (bifurcations) in Figs.
\ref{fig: MainResults}(I) and (II). The light purple circles indicate
the transition points that separate the stable branches with purely
real wavenumbers from the unstable branches with complex-valued wavenumbers.
Some features of these critical points have been previously explored
in \cite{rouhi2021exceptional,figotin2021exceptional}. Lastly, we
calculate the gain versus frequency diagram for the TWT using the
developed theoretical model, shown by the solid orange curve in Fig.
\ref{fig: MainResults}(c), and compare with results from computationally
intensive 3D PIC simulations, shown by blue crosses, demonstrating
very good agreement. The camel-like hump curve on the gain diagram
in Fig. \ref{fig: MainResults}(c) has the same shape as the unstable
branch in Fig. \ref{fig: MainResults}(b). The excellent agreement
between our developed theoretical results and PIC simulated results
in Fig. \ref{fig: MainResults} demonstrates the accuracy of our method.
Furthermore, to demonstrate the importance of the extra correction
factor $\delta_{\mathrm{e}}$ in our model, we also calculated the
gain versus frequency curve in Fig. \ref{fig: MainResults}(c) without
taking into account the correction factor $\delta_{\mathrm{e}}$ (dotted
light blue curve). In the case without a correction factor $\delta_{\mathrm{e}}$,
the calculated results are unable to predict the gain to within approximately
1.5 dB at the high amplification frequencies around $26\:\mathrm{GHz}$.

\section{Equivalent Circuit Model of Cold SWS\label{sec:Equivalent-Circuit-Model}}

It is crucial to have a simple model that estimates the cold (i.e.,
without the e-beam) characteristics of the SWS, especially for evaluating
the operational bandwidth and interaction efficiency of TWTs. Here,
we present the cold equivalent circuit model and compare frequency-dependent
cold results, such as phase velocity, with those of full-wave eigenmode
simulations.
\begin{figure}[t]
\begin{centering}
\includegraphics[width=0.73\columnwidth]{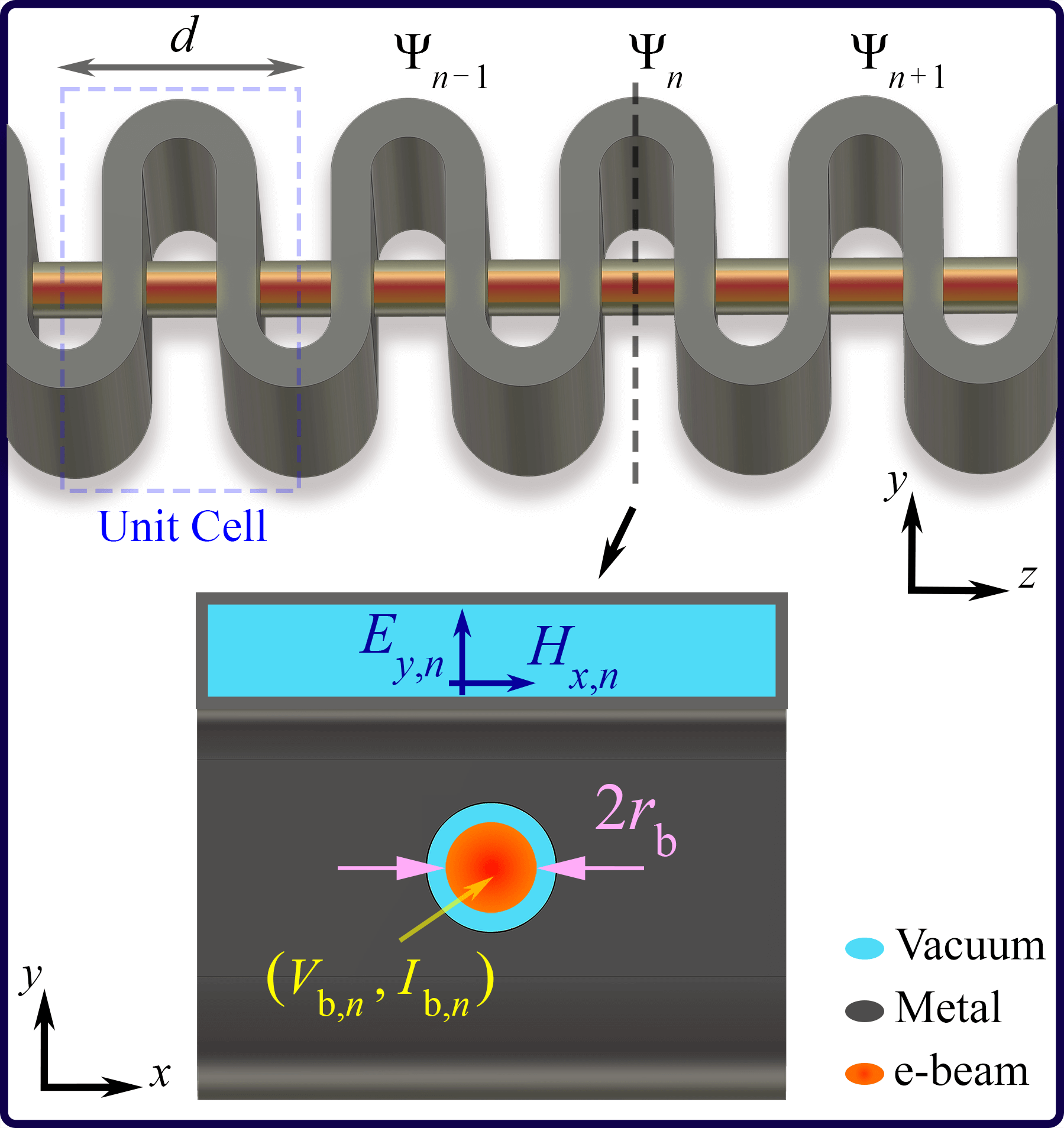}
\par\end{centering}
\centering{}\caption{Equivalent voltage and current at the input and output of each unit
cell and the corresponding TE10 electric and magnetic fields in the
cross section of the serpentine waveguide. We also show the equivalent
kinetic beam voltage and beam current pertinent to the two charge
waves.\label{fig: UnitCell}}
\end{figure}

A schematic design of an $E$-plane bend serpentine waveguide circuit
is shown in Fig. \ref{fig: Main}. It is assumed that only the fundamental
transverse-electric (TE) mode, i.e., TE10, propagates along the waveguide
with a rectangular cross section. In practice, reflections at the
junction with a bend cannot be completely avoided (segment J), and
we also need to take into account that the characteristic impedance
of the EM mode in the bend (segment B) is slightly different from
that of the EM mode in the straight segment; hence, the junction between
the two segments involves reactive fields \cite[Ch. 4]{marcuvitz1951waveguide,collin2001foundations}.
Note that the U-shaped bends in serpentine waveguides considered here
produce less reflection than the right angle bends commonly found
in folded waveguides \cite{ha1998theoretical,bhattacharjee2004folded,sumathy2013design}.
Additionally, reactive loading from the beam hole (segment H) can
affect device performance, depending on the hole's diameter. The effect
of both of these kinds of reflections is the creation of a stopband
that may limit the TWT's maximum operating frequency. In addition,
a band edge can also be a source of instability if an e-beam synchronizes
with it \cite{ha1998theoretical,hung2015absolute}. Thus, one must
carefully select a combination of beam tunnel radius and beam voltages
to avoid such an absolute instability at the $3\pi$ and $4\pi$ points
of the dispersion diagram, respectively.

The different segments of the serpentine waveguide are represented
in the center panel of Fig. \ref{fig: Main}, each with its own equivalent
TL circuit in the right panel. In this case, B, J, S, and H correspond
to the following parts of the unit cell: $E$-plane circular bend,
circular bend to straight waveguide junction, straight waveguide section,
and e-beam hole, respectively. By multiplying (cascading) the transfer
matrices of the individual segments we build the equivalent TL model
corresponding to the serpentine waveguide's unit cell which will be
further discussed in Subsection \ref{subsec:Cascaded-Circuit-Model}.
We use the equivalent representation in \cite{marcuvitz1951waveguide,felsen1994radiation}
that models propagation in a rectangular waveguide as a TL with equivalent
voltage and current. The discrete voltages and the currents that represent
the EM state in the phasor domain at different cross sections of the
waveguide are defined as $V_{n}=\sqrt{wb/2}E_{y,n}$ and $I_{n}=-\sqrt{wb/2}H_{x,n}$,
where $E_{y,n}$ and $H_{x,n}$ are the transverse electric and magnetic
fields of the TE10 mode calculated at the center of the rectangular
waveguide cross section as shown in Fig. \ref{fig: UnitCell}. The
equivalent voltage and current in the TLs are calculated at discrete
locations using the transfer matrix $\mathbf{\underline{T}}_{\mathrm{U}}$
as

\begin{equation}
\boldsymbol{\Psi}_{n}=\left[\begin{array}{c}
V_{n}\\
I_{n}
\end{array}\right],\;\boldsymbol{\Psi}_{n}=\mathbf{\underline{T}}_{\mathrm{U}}\boldsymbol{\Psi}_{n-1},
\end{equation}
where $V_{n-1}$ and $I_{n-1}$ are the equivalent voltage and current
\cite{marcuvitz1951waveguide} at the input port of the $n$th unit
cell and $V_{n}$ and $I_{n}$ are the equivalent voltage and current
at the output port of the $n$th unit cell as shown in Fig. \ref{fig: UnitCell}.

\subsection{Equivalent Matrix for Each Segment}

\subsubsection{Straight Waveguide (Segment S)\label{subsec:Straight-Waveguide}}

The straight rectangular waveguide segment of the unit cell is modeled
as a uniform TL of length $l$ with characteristic modal impedance
$Z_{0}=\eta_{0}/\sqrt{1-\left(\omega_{\mathrm{co}}/\omega\right)^{2}}$
of the fundamental TE10 mode, where $\eta_{0}=\sqrt{\mu_{0}/\varepsilon_{0}}$
is the wave impedance of free space, $\omega_{\mathrm{co}}=\pi c/w$
is the cutoff angular frequency, $w$ is the width of the rectangular
waveguide, and $\omega$ is the operating angular frequency. The phase
propagation constant of the TE10 mode is $\beta_{\mathrm{g,s}}=\sqrt{k_{0}^{2}-\left(\pi/w\right)}$,
where $k_{0}=2\pi/\lambda_{0}$, and $\lambda_{0}=2\pi c/\omega$
is wavelength in free space. The equivalent TL circuit representation
of the straight waveguide segment is shown in Fig. \ref{fig: Main}
(segment S), and the equivalent transfer matrix is

\begin{equation}
\mathbf{\underline{T}}_{\mathrm{S}}=\left[\begin{array}{rr}
\cos\left(\beta_{\mathrm{g,s}}l\right) & jZ_{0}\sin\left(\beta_{\mathrm{g,s}}l\right)\\
j\sin\left(\beta_{\mathrm{g,s}}l\right)/Z_{0} & \cos\left(\beta_{\mathrm{g,s}}l\right)
\end{array}\right].
\end{equation}

\subsubsection{Circular Bend to Straight Waveguide Junction (Segment J)}

The junction between the straight waveguide and the $E$-plane bend
is represented by the equivalent circuit in Fig. \ref{fig: Main}
(segment J) with equivalent lumped reactance \cite[Sec. 5.34]{marcuvitz1951waveguide}

\begin{equation}
X=Z_{0}\left(\frac{32}{\pi^{7}}\left(\frac{2\pi b}{\lambda_{\mathrm{g,s}}}\right)^{3}\left(\frac{b}{R}\right)^{2}\sum_{n=1,3,\ldots}^{\infty}\frac{1}{n^{7}}\sqrt{1-\left(\frac{2b}{n\lambda_{\mathrm{g,s}}}\right)^{2}}\right),
\end{equation}
where $R$ is the mean radius of the bend, and $\lambda_{\mathrm{g,s}}=2\pi/\beta_{\mathrm{g,s}}=\lambda_{0}/\sqrt{1-\left(\omega_{\mathrm{co}}/\omega\right)^{2}}$
is the guided wavelength. The equivalent transfer matrix for the junction
is

\begin{equation}
\mathbf{\underline{T}}_{\mathrm{J}}=\left[\begin{array}{rr}
1 & -jX\\
0 & 1
\end{array}\right].
\end{equation}

\subsubsection{$E$-plane Circular Bend (Segment B)}

An equivalent TL circuit for the quarter $E$-plane bend is given
in Fig. \ref{fig: Main} (segment B). Here, $\pi R/2$ is the mean
length of the $E$-plane bend and the length of the equivalent TL.
The modified characteristic impedance for the fundamental propagating
mode in the bend is \cite[Sec. 5.34]{marcuvitz1951waveguide}

\begin{equation}
Z_{0,\mathrm{b}}=Z_{0}\left(1+\frac{1}{12}\left(\frac{b}{R}\right)^{2}\left[\frac{1}{2}-\frac{1}{5}\left(\frac{2\pi b}{\lambda_{\mathrm{g,s}}}\right)^{2}\right]\right).\label{eq:Z_TE10,b}
\end{equation}
In addition, the circular bend is considered as a uniform angular
waveguide with a guided wavelength of

\begin{equation}
\lambda_{\mathrm{g,b}}\simeq\lambda_{\mathrm{g,s}}\left(1-\frac{1}{12}\left(\frac{b}{R}\right)^{2}\left[-\frac{1}{2}+\frac{1}{5}\left(\frac{2\pi b}{\lambda_{\mathrm{g,s}}}\right)^{2}-\ldots\right]\right),\label{eq:Lambda_g,b}
\end{equation}
for the fundamental mode. As a result, in the wavelength range $2b/\lambda_{\mathrm{g,s}}<1$
\cite[Sec. 5.34]{marcuvitz1951waveguide}, the TL matrix for the circular
bend segment is

\begin{equation}
\mathbf{\underline{T}}_{\mathrm{B}}=\left[\begin{array}{rr}
\cos\left(\frac{\pi^{2}R}{\lambda_{\mathrm{g,b}}}\right) & jZ_{0,\mathrm{b}}\sin\left(\frac{\pi^{2}R}{\lambda_{\mathrm{g,b}}}\right)\\
j\sin\left(\frac{\pi^{2}R}{\lambda_{\mathrm{g,b}}}\right)/Z_{0,\mathrm{b}} & \cos\left(\frac{\pi^{2}R}{\lambda_{\mathrm{g,b}}}\right)
\end{array}\right].
\end{equation}

\subsubsection{Beam Tunnel Hole}

The radius of the beam hole can slightly affect the phase velocity,
dispersion and cutoff frequency of the EM mode in the SWS \cite{liu2000study,li2013dispersion}.
A wide beam tunnel will add significant periodic reactive loading
to the SWS and introduce a stopband at the $3\pi$ point of the modal
dispersion diagram, and the larger beam tunnel radius results in a
larger stopband \cite{nguyen2014design}. On the other hand, a wide
beam tunnel permits higher beam currents since the beam radius can
be larger with the same current density, resulting in higher d.c.
beam power and output RF power at saturation \cite{meyne2017simulation}.
However, an e-beam of a very small radius (with the same d.c. beam
current) will experience strong Coulomb repulsion between electrons,
and it is unrealistic to apply an intense magnetic field to confine
an e-beam with a small radius and high current density \cite{nguyen2014design}.
Also, it is desirable to have an e-beam with a lower accelerating
voltage and a higher current, resulting in a higher gain. Therefore,
it is necessary to trade off beam tunnel size, current density, and
beam radius to optimize TWT properties, such as linear gain and efficiency.

A general and accurate circuit to model the beam tunnel hole that
can be used in all cases has not been developed yet. In \cite{choi1995folded},
the authors modeled the circular hole as a shunt reactance, where
the value depends on rectangular waveguide width and height and beam
tunnel diameter. Also, in \cite{booske2005accurate}, a circuit model
of the beam tunnel hole based on the modification of the model for
different tunnel radii in \cite{marcuvitz1951waveguide} was presented.
The reference structure is a circular waveguide connected orthogonally
to the broad wall of a rectangular waveguide through a small aperture.
The difference between the reference structure in \cite{marcuvitz1951waveguide}
and the structure to be modeled is that the cylindrical tunnel is
represented as a stub whose diameter equals the aperture diameter
and is below the cutoff for propagation and there are two of these
stubs present. By assuming that the hole radius is electrically small
(i.e., much smaller than the guided wavelength), we can often neglect
the effect of holes and model this section as a simple straight rectangular
waveguide as described in Subsection \ref{subsec:Straight-Waveguide}.
This approximation leads to acceptable results and more investigation
for a specific example is provided in Appendix \ref{app:Non-cross-Section-Interaction}.
In addition, several papers designed serpentine waveguide TWTs without
considering the effect of the beam tunnel hole, and some papers used
the straight waveguide model for it, including \cite{choi1995folded,liu1995folded,na2002analysis,han2003design}.

In TWTs designed for millimeter waves and even higher frequencies,
the e-beam tunnel is often enlarged to achieve higher transmission
rates, thereby causing a bandgap at the $3\pi$ point. For large beam
tunnel dimensions, one could obtain the $S$-parameters of the straight
segment with non-negligible tunnel loading via full-wave simulations.
The numerically obtained $S$-parameters can then be converted into
the transmission matrix $\mathbf{\underline{T}}_{\mathrm{H}}$ and
used in our model. However, if circuit models for the segment with
a large beam tunnel become available, one could also include them
in the present formulation.

\subsection{Cascaded Circuit Model\label{subsec:Cascaded-Circuit-Model}}

The basic SWS segments shown in Fig. \ref{fig: Main} are represented
by equivalent TL segments, each with an equivalent transfer matrix
as discussed above. The transfer matrices for the lossless circuit
segments are cascaded to arrive at the transfer matrix of the unit
cell represented as

\begin{equation}
\begin{array}{c}
\mathbf{\underline{T}}_{\mathrm{U}}=\left[\begin{array}{cc}
T_{11} & T_{12}\\
T_{21} & T_{22}
\end{array}\right]=\left(\mathbf{\underline{T}}_{\mathrm{U/2}}\right)^{2}.\end{array}\label{eq:Equivalent Matrix}
\end{equation}
For convenience we use the half unit cell transfer matrix defined
as

\begin{equation}
\mathbf{\underline{T}}_{\mathrm{U/2}}=\left(\mathbf{\underline{T}}_{\mathrm{B}}\mathbf{\underline{T}}_{\mathrm{J}}\mathbf{\underline{T}}_{\mathrm{S}}\mathbf{\underline{T}}_{\mathrm{H}}\mathbf{\underline{T}}_{\mathrm{S}}\mathbf{\underline{T}}_{\mathrm{J}}\mathbf{\underline{T}}_{\mathrm{B}}\right).
\end{equation}
Using our unit cell transfer matrix $\mathbf{\underline{T}}_{\mathrm{U}}$,
we find solutions for the state vector, $\boldsymbol{\Psi}=\left[V,I\right]^{\mathrm{T}}$,
that satisfies

\begin{equation}
\mathbf{\underline{T}}_{\mathrm{U}}\boldsymbol{\Psi}=e^{-\mathit{j}\beta_{\mathrm{c,0}}\mathit{d}}\boldsymbol{\Psi},
\end{equation}
where $d$ is unit cell period and $\mathit{\beta_{\mathrm{c,0}}}$
is the wavenumber of the fundamental spatial harmonic. Solving the
eigenvalue problem,

\begin{equation}
\mathrm{det\left(\mathbf{\underline{T}}_{\mathrm{U}}-{\it e}^{-\mathit{j}\mathit{\beta_{\mathrm{c,0}}}\mathit{d}}\mathbf{\underline{I}}\right)}=0,
\end{equation}
for $\beta_{\mathrm{c,0}}$, yields the Bloch wavenumbers of the cold
EM modes allowed in the SWS, where $\mathbf{\underline{I}}$ is the
$2\times2$ identity matrix. Then, the propagation constants for the
$m$th spatial harmonic is

\begin{equation}
\beta_{\mathrm{c},m}=\beta_{\mathrm{c,0}}+\frac{2m\pi}{d},\quad m=0,\pm1,\pm2,\ldots
\end{equation}
The phase velocity of the spatial harmonic of the cold mode is calculated
as $v_{\mathrm{c},m}=\omega/\beta_{\mathrm{c},m}$. Based on the definition
of the state vector at the beginning of each unit cell, the characteristic
Bloch impedance of the fundamental guided mode is calculated as

\begin{equation}
Z_{\mathrm{c}}=\frac{V}{I}=\frac{T_{12}}{e^{-\mathit{j}\beta_{\mathrm{c},0}\mathit{d}}-T_{11}}=\frac{e^{-\mathit{j}\beta_{\mathrm{c},0}\mathit{d}}-T_{22}}{T_{21}}.\label{eq: ImpCascade}
\end{equation}
Note that the characteristic Bloch impedance depends on where the
section separating unit cells is defined, and if we substitute $\beta_{\mathrm{c},0}$
for $\beta_{\mathrm{c},m}$, the result does not change.

\subsection{Equivalent Uniform TL Model\label{subsec:Equivalent-Uniform-TL}}

Each EM mode is comprised of a fundamental Bloch wavenumber $\beta_{\mathrm{c,0}}$
and all its spatial harmonics $\beta_{\mathrm{c},m}$. However, the
Pierce model \cite{pierce1947theory,pierce1949new,pierce1950Book,pierce1951waves}
is based on the assumption that the SWS can be considered as a uniform
TL supporting a single mode with wavenumber $\beta_{\mathrm{c}}$
that is velocity-synchronized with the e-beam, which is discussed
here. To highlight this view, we impose that the cascaded matrix $\mathbf{\underline{T}}_{\mathrm{U}}$
in (\ref{eq:Equivalent Matrix}) should be equal to the transfer matrix
of an equivalent \textit{uniform} single TL, as was done also in \cite{choi1995folded,booske2005accurate},

\begin{equation}
\begin{array}{c}
\mathbf{\underline{T}}_{\mathrm{Uni}}=\left[\begin{array}{rr}
\cos\left(\beta_{\mathrm{c,0}}d\right) & jZ_{\mathrm{c}}\sin\left(\beta_{\mathrm{c,0}}d\right)\\
j\sin\left(\beta_{\mathrm{c,0}}d\right)/Z_{\mathrm{c}} & \cos\left(\beta_{\mathrm{c,0}}d\right)
\end{array}\right].\end{array}\label{eq:Equivalent Uniform}
\end{equation}
Then, we impose $\mathbf{\underline{T}}_{\mathrm{Uni}}=\mathbf{\underline{T}}_{\mathrm{U}}$,
where $\mathbf{\underline{T}}_{\mathrm{U}}$ is calculated from the
cascaded circuit equivalent model of each segment as explained in
the previous subsection, and we obtain the elements of $\mathbf{\underline{T}}_{\mathrm{Uni}}$
for $\beta_{\mathrm{c,0}}d$, which is the effective phase shift per
unit cell of the fundamental spatial harmonic. As a result, the propagation
constants for all spatial harmonics are \cite[Sec. 4.5.1]{carter2018microwave}

\begin{equation}
\beta_{\mathrm{c},m}=\frac{\cos^{-1}\left(T_{\mathrm{11}}\right)}{d}+\frac{2m\pi}{d},\quad m=0,\pm1,\pm2,\ldots,
\end{equation}
where $m$ denotes the harmonic number. In serpentine waveguide TWTs
usually the first spatial harmonic ($m=1$) is synchronized with the
e-beam. The phase velocity corresponding to the $m$th spatial harmonic
is $v_{\mathrm{c},m}=\omega/\beta_{\mathrm{c},m}$. The second and
third elements in the equivalent transmission matrix $\mathbf{\underline{T}}_{\mathrm{Uni}}$
are used to calculate the characteristic impedance of the equivalent
uniform TL as $Z_{\mathrm{c}}=\sqrt{T_{\mathrm{12}}/T_{\mathrm{21}}}$
\cite[Sec. 4.5.1]{carter2018microwave}. Also, by imposing $\mathbf{\underline{T}}_{\mathrm{Uni}}=\mathbf{\underline{T}}_{\mathrm{U}}$
to (\ref{eq:Equivalent Matrix}) and using the reciprocity property
of the transfer matrix, the latter equation for characteristic impedance
will be equivalent to (\ref{eq: ImpCascade}).

\begin{figure}[t]
\begin{centering}
\includegraphics[width=1\columnwidth]{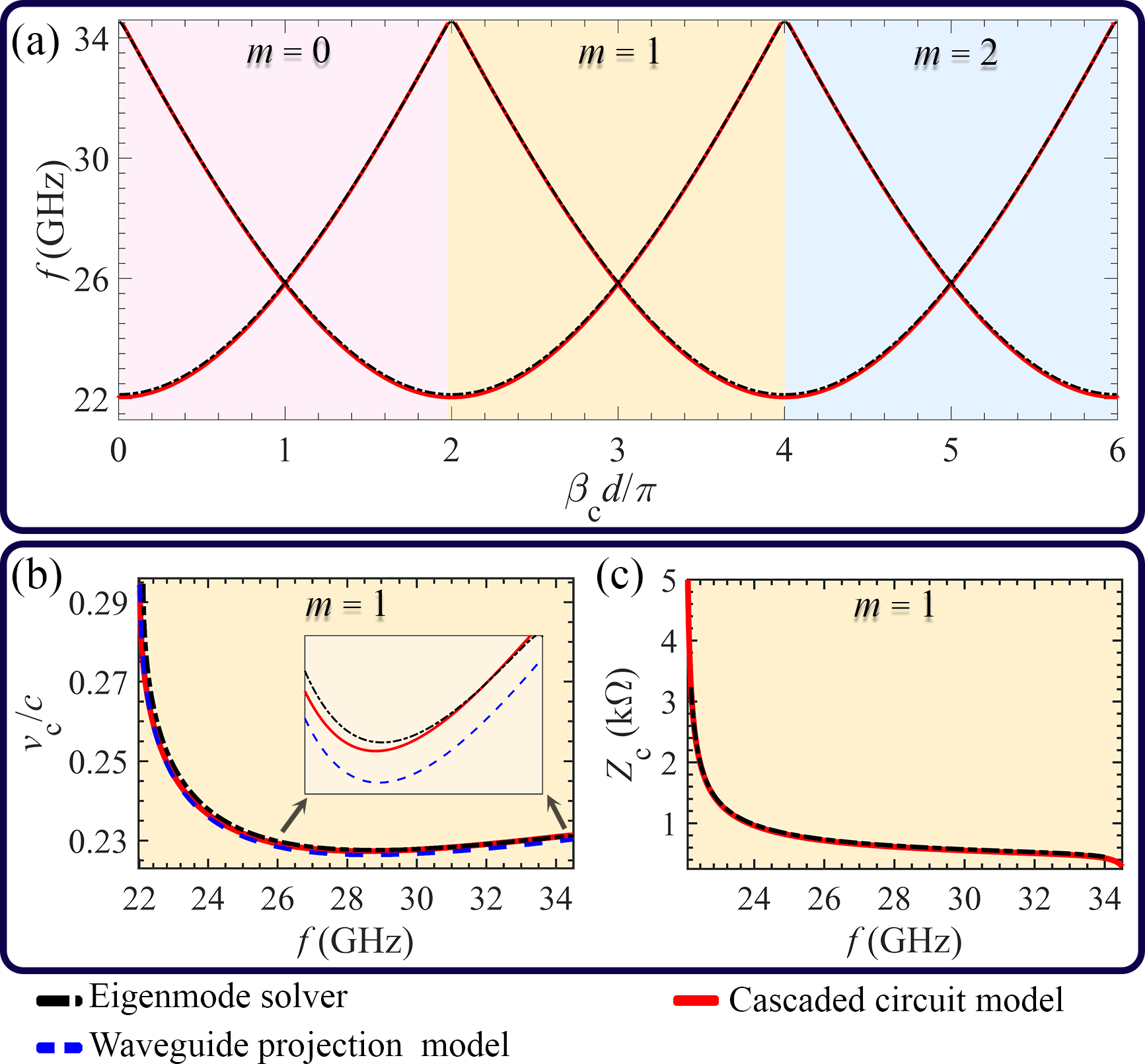}
\par\end{centering}
\centering{}\caption{Cold simulation results for the serpentine waveguide using theoretical
and simulation methods. (a) Modal dispersion curves for three spatial
harmonics ($m=0,1$ and $2$) by employing the full-wave eigenmode
solver (dashed black curves) and cascaded circuit model (solid red
curve). (b) Normalized phase velocity for the first spatial harmonic
($m=1$) by using the eigenmode solver (dashed black curves), cascaded
circuit model (solid red curve) and waveguide projection model (dashed
blue curves). Also, the zoomed-in version of normalized phase velocity
in the frequency range from $26\:\mathrm{GHz}$ to $34.5\:\mathrm{GHz}$
is shown to demonstrate the superior accuracy of the cascaded circuit
model compared to the waveguide projection model. (c) Characteristic
Bloch impedance for the first spatial harmonic ($m=1$) calculated
using the cascaded circuit model (solid red curve), compared with
that from full-wave simulation (dashed black curve).\label{fig: ColdResults}}
\end{figure}

\subsection{Waveguide Projection Model (Without Considering the Junction and
Bend Effect)\label{subsec:Waveguide-Projection-Model}}

The guided wavenumbers $\beta_{\mathrm{c},m}$ can also be approximated
by considering the SWS as a straightened version of the serpentine
waveguide. In this simple view, the effect of the junction between
straight and bend sections is ignored and we assume that the TE10
propagation constant in the curved segments is the same as in the
straight segments. The on-axis phase shift per pitch for the $m$th
spatial harmonics is $\beta_{\mathrm{c},m}d=\theta+2m\pi$, where
$\beta_{\mathrm{c},m}$ is the \textit{effective} on-axis propagation
constant, $\theta=\beta_{\mathrm{g,s}}L$ is phase delay per pitch
of EM wave, and $L=2\left(\pi R+h\right)$ is defined as the distance
traveled by the wave per pitch. The phase velocity of $m$th spatial
harmonics is expressed by \cite{zheng2009parametric}

\begin{equation}
v_{\mathrm{c},m}=\frac{\omega}{\beta_{\mathrm{c},m}}=\frac{\omega d}{\beta_{\mathrm{g,s}}L+2\pi m}.\label{eq:SimpleModel}
\end{equation}
The derivation of (\ref{eq:SimpleModel}) assumes that the bends do
not present significant mismatches to the wave. In practice, both
the bends and the beam holes introduce small mismatches that may cause
stopbands where the dispersion curves of spatial harmonics cross.
These effects are ignored in this simplified model.

\section{Validation of Equivalent Circuit Model\label{sec:Validation-of-Equivalent}}

The cold SWS characteristics for a specific design are shown via the
three theoretical models discussed in the previous section, compared
with simulations performed using the CST Studio Suite eigenmode solver.
Figure \ref{fig: Main} shows the model of a typical serpentine waveguide
with a cylindrical beam tunnel, where the geometric parameters $w$,
$b$, $d$, $h$, and $r_{\mathrm{c}}$ represent the dimensions of
wide side, narrow side, full period, straight waveguide wall, and
radius of beam tunnel, respectively. The parameter values for a specific
design are listed in Table \ref{tab:Parameters}.
\begin{figure}[t]
\begin{centering}
\includegraphics[width=1\columnwidth]{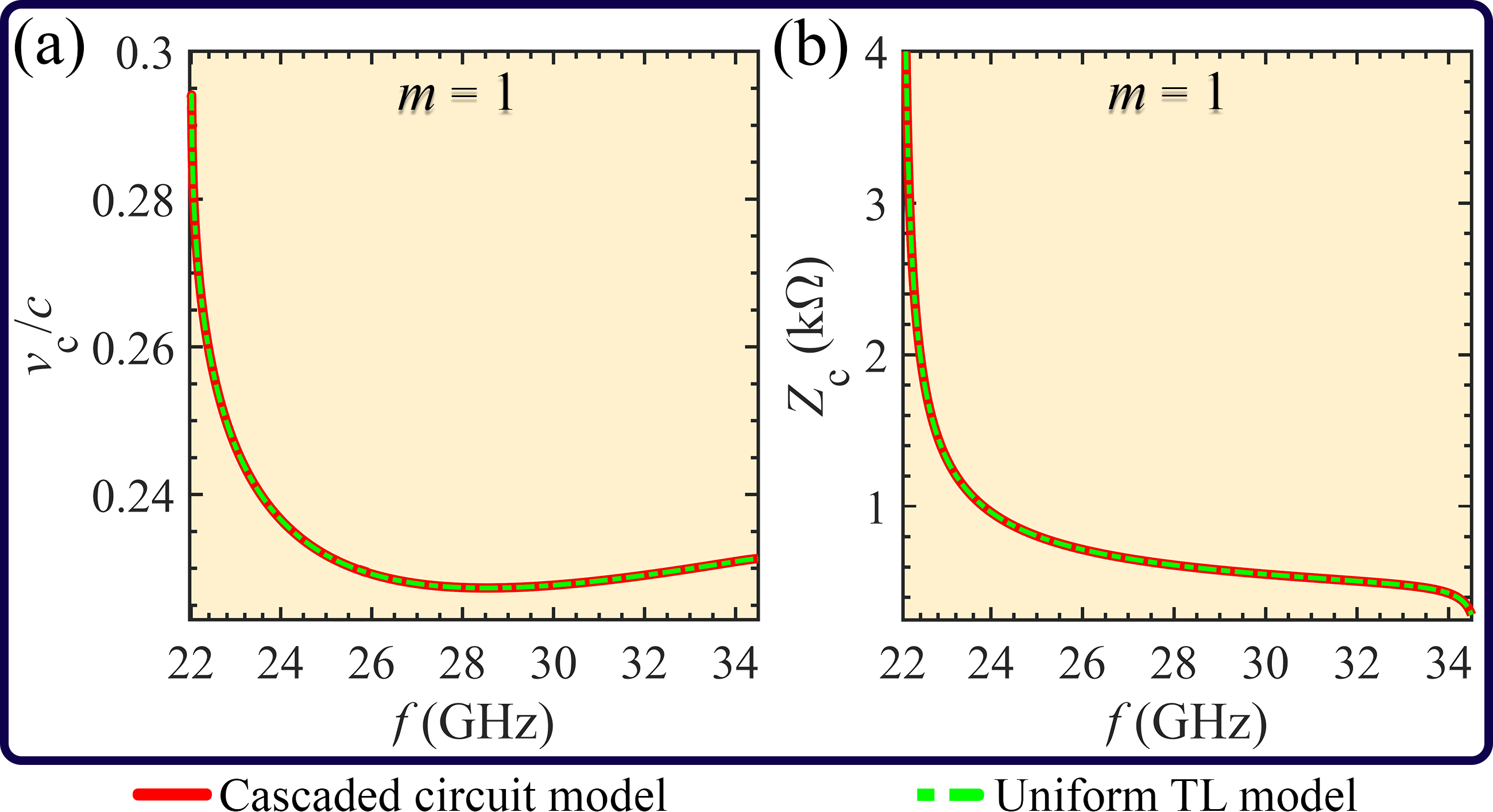}
\par\end{centering}
\centering{}\caption{(a) Normalized phase velocity and (b) characteristic Bloch impedance
of the first spatial harmonic ($m=1$) calculated by using the cascaded
circuit model (solid red curve) described in Subsection \ref{subsec:Cascaded-Circuit-Model}
and the uniform TL model (dashed green curve) described in Subsection
\ref{subsec:Equivalent-Uniform-TL}.\label{fig: ColdUni}}
\end{figure}
\begin{table}[tbh]
\caption{Designed structural parameters for the serpentine waveguide SWS.\label{tab:Parameters}}

\centering{}%
\begin{tabular}{|c|c|c|}
\hline 
Description & Parameter & Length (mm)\tabularnewline
\hline 
\hline 
The width of rectangular waveguide & $w$ & $6.8$\tabularnewline
\hline 
The height of rectangular waveguide & $b$ & $0.7$\tabularnewline
\hline 
The full period length & $d$ & $4$\tabularnewline
\hline 
The whole straight waveguide length & $h$ & $2.5$\tabularnewline
\hline 
The radius of beam tunnel & $r_{\mathrm{c}}$ & $0.5$\tabularnewline
\hline 
\end{tabular}
\end{table}

Figure \ref{fig: ColdResults} shows the wavenumber, phase velocity
and characteristic impedance of the EM modes in the cold serpentine
waveguide obtained using theoretical and simulation methods. Figure
\ref{fig: ColdResults}(a) shows the wavenumber dispersion diagram
of the modes in the serpentine waveguide, showing three spatial harmonics,
obtained by varying the phase between periodic boundaries. The simulated
results based on the CST Studio Suite eigenmode solver (dashed black
curves) are in excellent agreement with the theoretical dispersion
diagram calculated by the cascaded circuit model (solid red curves)
discussed in Subsection \ref{subsec:Cascaded-Circuit-Model}. The
cutoff frequency of the designed serpentine waveguide is around $f_{\mathrm{c}}=22.15\:\mathrm{GHz}$.
Then, the normalized phase velocity corresponding to the first spatial
harmonic ($m=1$) as a function of frequency ranging from $22.15\:\mathrm{GHz}$
to $34.5\:\mathrm{GHz}$ is plotted in Fig. \ref{fig: ColdResults}(b).
There is excellent agreement between the results provided by the eigenmode
solver (dashed black curve), cascaded circuit model (solid red curve)
in Subsection \ref{subsec:Cascaded-Circuit-Model}, and waveguide
projection model (dashed blue curve) in Subsection \ref{subsec:Waveguide-Projection-Model}.
As a general observation, the cascaded circuit model is more accurate
than the waveguide projection model because it accounts for the mismatches
due to circular bends and junctions. The characteristic Bloch impedance
of the serpentine waveguide SWS using the cascaded circuit model (solid
red curve) in Section \ref{subsec:Cascaded-Circuit-Model}, compared
to that from full-wave simulation (dashed black curves), is shown
in Fig. \ref{fig: ColdResults}(c). The characteristic Bloch impedance
from full-wave simulation is calculated as $Z_{\mathrm{c}}=-E_{y}/H_{x}=V/I$,
by using field monitors.

In order to demonstrate that the serpentine waveguide can be modeled
by a single straight uniform TL (Section \ref{subsec:Equivalent-Uniform-TL}),
we compare the results based on the uniform TL model with the cascaded
circuit model (Section \ref{subsec:Cascaded-Circuit-Model}). The
calculated phase velocity and characteristic impedance results for
the first spatial harmonic in both cases are shown in Fig. \ref{fig: ColdUni}(a)
and (b), and we observe excellent agreement between these two theoretical
methods. Also, previous studies, such as \cite{booske2005accurate},
utilized the uniform TL model that is very similar to what is discussed
in this paper for the cold case. In contrast, in this paper we also
develop a model for finding the ``hot eigenmodes'' dispersion of
the device and the TWT gain. The accurate calculation of the characteristic
parameters of the cold structure, i.e., $Z_{\mathrm{c}}$ and $v_{\mathrm{c}}$,
has a vital role in our model. To reinforce this point, we note that
one of the conclusions of \cite{booske2005accurate} is that accurate
determination of the small-signal gain in a serpentine waveguide TWT
amplifier requires a precise evaluation of the phase velocity to within
$0.5\%$ and the interaction impedance within $10\%$ of the actual
parameters found by time-consuming full-wave eigenmode simulations.
The calculated gain is very sensitive to these parameters, and requires
correct phase velocity and interaction impedance specification. Sensitivity
studies in \cite{booske2005accurate} indicate that variations in
the phase velocity of $0.5\%$ can result in $8\:\mathrm{dB}$ of
variation in the predicted small-signal gain, while a $10\%$ variation
in the interaction impedance can result in a $5\:\mathrm{dB}$ change
in the predicted small-signal gain of the specific design.

\section{Interaction Impedance\label{sec:Pierce-Impedance}}

In order to predict the performance of a TWT, one needs to determine
the interaction (Pierce) impedance of the serpentine waveguide because
amplifier gain is proportional to the cubic root of this parameter
\cite{pierce1950Book}. The interaction impedance is a measure of
how much the on-axis electric field can velocity modulate electrons
for a given EM power propagating along the length of the structure
\cite[Ch. 10]{gewartowski1965principles}. In the ideal case, the
e-beam is assumed to be very narrow. From Pierce theory, the interaction
impedance for a thin beam is defined for a specific spatial harmonic
$m$ as \cite[Ch. 10]{gewartowski1965principles}

\begin{equation}
Z_{\mathrm{P},m}\left(\beta_{\mathrm{c,}m}\right)=\frac{\left|E_{z,m}\left(\beta_{\mathrm{c,}m}\right)\right|^{2}}{2\beta_{\mathrm{c,}m}^{2}P},\label{eq:Zpierce}
\end{equation}

\noindent where $\left|E_{z,m}\left(\beta_{\mathrm{c,}m}\right)\right|$
is the magnitude of the axial electric field phasor along the center
of the cold SWS where the e-beam will be introduced, for a given phase
constant and spatial harmonic $m$, and $P$ is the time-average power
flux through the SWS at the given phase propagation constant $\beta_{\mathrm{c,}m}$
\cite{gewartowski1965principles}. The quantity $\left|E_{z,m}\right|$
is the weight of the $m$th Floquet-Bloch spatial harmonics of the
axial field decomposition $E_{z}\left(z,\beta_{\mathrm{c}}\right)=\sum_{m=-\infty}^{\infty}E_{z,m}\left(\beta_{\mathrm{c}}\right)e^{-j\beta_{\mathrm{c,}m}z}$.
It is calculated by numerically obtaining the phasor of the axial
electric field $E_{z}\left(z,\beta_{\mathrm{c}}\right)$ of the cold
serpentine waveguide with beam tunnel using full-wave eigenmode simulations,
followed by performing the Fourier transform in space
\begin{equation}
E_{z,m}\left(\beta_{\mathrm{c}}\right)=\frac{1}{d}\intop_{0}^{d}E_{z}\left(z,\beta_{\mathrm{c}}\right)e^{j\beta_{\mathrm{c,}m}z}dz.\label{eq:ZpierceCoef}
\end{equation}

In addition, the time average power flux is simply calculated as $P=W_{\mathrm{t}}v_{\mathrm{g}}/d$
\cite{sharma2014design}, where $W_{\mathrm{t}}$ is the total EM
energy of the wave stored in a unit cell and $v_{\mathrm{g}}=d\omega/d\beta_{\mathrm{c}}$
is the group velocity. For a serpentine waveguide, the interaction
impedance is typically evaluated within the first Brillouin zone (i.e.,
$m=1$), where the interaction occurs. Additionally, the e-beam diameter
also influences the interaction impedance. For beam cross sections
and beam tunnel diameters that are not infinitesimally thin, the longitudinal
electric field and the interaction impedance within the beam tunnel
can vary over the beam cross section area, becoming larger near the
edges of the tunnel. As a consequence, the additional correction factor
(average factor) should be considered in calculating the interaction
impedance by taking into account the variation of the electric field
within the interaction area (interaction gap) \cite{sudhamani2017investigation}.
Additional analysis of the variation of the electric field distribution
in the interaction area for the specific example can be found in Appendix
\ref{app:Non-uniform-Interaction}. Therefore, a modified or ``effective
interaction impedance'' corresponding to each spatial harmonic considering
the nonuniform electric field distribution in the interaction area
is given by

\begin{equation}
Z_{\mathrm{P},\mathrm{e},m}=\left(1+\delta_{\mathrm{e}}\right)^{2}Z_{\mathrm{P},m},\label{eq:ZPierceCorrectDelta}
\end{equation}
where, $\delta_{\mathrm{e}}>0$ is the correction factor. The value
of the correction factor $\delta_{\mathrm{e}}$ can be found either
by (i) averaging the EM axial field over the beam cross section, or
(ii) by matching the maximum value of the theoretical and PIC-simulated
gain at the synchronization frequency.
\begin{figure}[t]
\begin{centering}
\includegraphics[width=0.85\columnwidth]{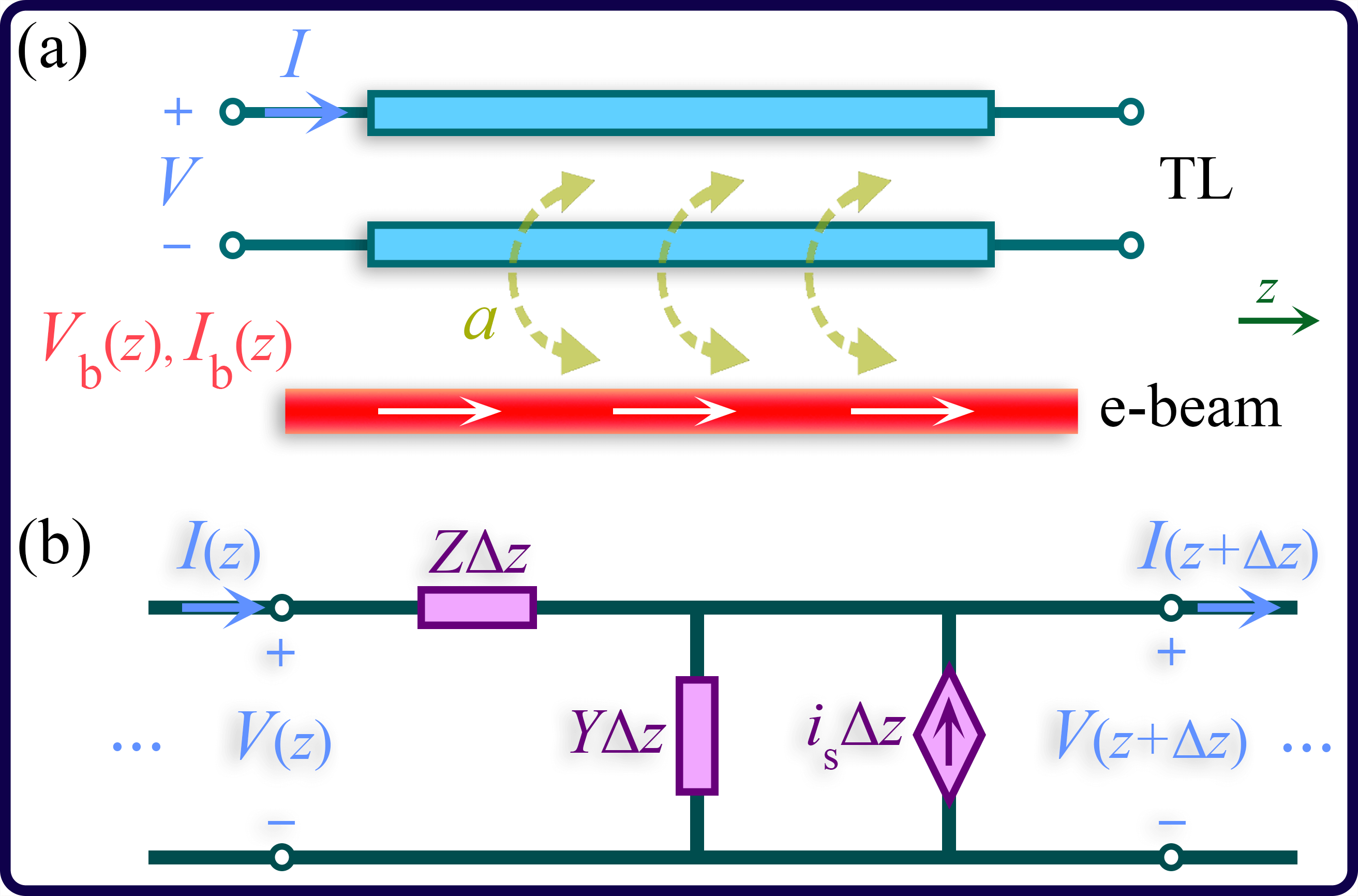}
\par\end{centering}
\centering{}\caption{(a) Schematic of the equivalent TL coupled to the e-beam used to study
the beam-EM wave interaction in the serpentine waveguide TWT. (b)
Equivalent TL circuit showing the per-unit-length impedance, admittance
and current generator $i_{\mathrm{s}}=-a\partial_{z}I_{\mathrm{b}}$
that represents the effect of the e-beam on the TL.\label{fig: TLHot}}
\end{figure}

In this paper, to compute the interaction impedance $Z_{\mathrm{P},m}$,
we use the eigenmode solver of CST Studio Suite to calculate $E_{z}\left(z,\beta_{\mathrm{c}}\right)$
over $z$ for different $\beta_{\mathrm{c}}$. Then, we transform
the electric field $E_{z}\left(z,\beta_{\mathrm{c}}\right)$ by (\ref{eq:ZpierceCoef})
and calculate the interaction impedance by (\ref{eq:Zpierce}). The
group velocity $v_{\mathrm{g}}=d\omega/d\beta_{\mathrm{c}}$ is determined
directly from the dispersion diagram by using numerical differentiation.
The EM energy simulated within a single unit cell between periodic
boundaries in the eigenmode solver is always $1~\mathrm{Joule}$.

\section{E-beam and EM Wave Interaction\label{sec:e-beam-and-EM}}

The classical small-signal theory by J. R. Pierce is one of the most
famous approaches used for TWT modeling and design \cite{pierce1947theory,pierce1949new,pierce1950Book,pierce1951waves}.
Our implementation based on the generalization of Pierce's theory
is summarized here following our previous work \cite{rouhi2021exceptional}.
We follow the linearized equations that describe the space-charge
wave as originally presented by Pierce.

The equivalent model for the TWT describes the EM wave traveling in
a serpentine waveguide interacting with an e-beam flowing in the $z$
direction as shown schematically in Fig. \ref{fig: TLHot}. The electrons
have an average velocity and linear charge density of $u_{0}$ and
$\rho_{0}$, respectively. The e-beam has an average current $I_{0}=-\rho_{0}u_{0}$
in the $-z$ direction and an equivalent kinetic d.c. voltage as $V_{0}\approx u_{0}^{2}/\left(2\eta\right)$
for non-relativistic beams (assuming that thermal initial velocity
of the electron is neglected) or $V_{0}=\left[\left(1-\left(u_{0}/c\right)^{2}\right)^{-1/2}-1\right]c^{2}/\eta$
for relativistic beams, where $c$ is the speed of light in a vacuum,
$\eta=e/m=1.758820\times10^{11}\:\mathrm{C/Kg}$ is the charge-to-mass
ratio of the electron with charge $-e$ and rest mass $m$ \cite[Ch. 3]{gilmour1994principles}. The model we developed is based on a non-relativistic beam. 
The small-signal modulations in the charge velocity $u_{\mathrm{b}}$
and charge density $\rho_{\mathrm{b}}$, describe the \textquotedblleft space-charge
wave\textquotedblright . The a.c. equivalent beam current and kinetic
voltage are given by $i_{\mathrm{b}}=u_{\mathrm{b}}\rho_{0}+u_{0}\rho_{\mathrm{b}}$
and $v_{\mathrm{b}}=u_{\mathrm{b}}u_{0}/\eta$, where we have kept
only the linear terms based on the small-signal approximation \cite{pierce1951waves}.
We implicitly assume a time dependence of $\exp\left(j\omega t\right)$,
so the a.c. space-charge wave modulating the e-beam is described in
the phasor domain with $V_{\mathrm{b}}\left(z\right)$ and $I_{\mathrm{b}}\left(z\right)$,
as

\begin{equation}
\frac{d}{dz}V_{\mathrm{b}}=-j\beta_{0}V_{\mathrm{b}}-aZI-j\frac{I_{\mathrm{b}}}{A\varepsilon_{0}\omega},\label{eq: V_b}
\end{equation}

\begin{equation}
\frac{d}{dz}I_{\mathrm{b}}=-jgV_{\mathrm{b}}-j\beta_{0}I_{\mathrm{b}},\label{eq: I_b}
\end{equation}
where $\beta_{0}=\omega/u_{0}$ is the space-charge wave equivalent
phase constant (when neglecting plasma frequency effects), $g=I_{0}\beta_{0}/\left(2V_{0}\right)$,
$Z$ is the equivalent TL distributed series impedance, and $I\left(z\right)$
is the equivalent TL current. The term $E_{\mathrm{w}}=aZI$ is the
longitudinal electric field of the EM mode propagation in the SWS,
affecting the bunching of the e-beam. In addition, the coefficient
$a$ represents a coupling strength that describes how the e-beam
couples to the TL, already introduced in \cite{tamma2014extension,rouhi2021exceptional,rouhi2023modeling}
and \cite[Ch. 3]{figotin2013multi,figotin2020analytic} and investigated
in more detail in Appendix \ref{app:Coupling-Strength-Coefficient}.
Also, the term $E_{\mathrm{p}}=jI_{\mathrm{b}}/\left(A\varepsilon_{0}\omega\right)$
is the longitudinal electric field term arising from the nonuniform
charge density that causes the so-called ``debunching'' \cite[Ch. 10]{gewartowski1965principles},
where $A$ is the e-beam cross sectional area, and $\varepsilon_{0}$
is vacuum permittivity. This field is responsible of the repulsive
forces in a dense beam of charged particles. Therefore, $E_{z}=E_{\mathrm{w}}+E_{\mathrm{p}}$
is the total longitudinal $z$-polarized electric field component
in the hot structure (when also the e-beam is present) that modulates
the velocity and bunching of the electrons. In serpentine waveguide
TWTs, the beam-EM wave interaction occurs in the first spatial harmonic
($m=1$), so in this section we drop the subscript harmonic index
\textit{$m$} for simplicity. The telegrapher\textquoteright s equations,

\begin{equation}
\frac{d}{dz}V=-ZI,\label{eq: TelegV}
\end{equation}

\begin{equation}
\frac{d}{dz}I=-YV-a\frac{d}{dz}I_{\mathrm{b}},\label{eq: TelegI}
\end{equation}
describe the modal propagation in the SWS of the EM mode synchronizing
with the e-beam in terms of equivalent TL voltage $V\left(z\right)$
and current $I\left(z\right)$ phasors, based on the equivalent TL
model shown in Fig. \ref{fig: TLHot}(b). Figure \ref{fig: TLHot}(b)
shows the distributed per-unit-length series impedance $Z$ and shunt
admittance $Y$ as well as the term $i_{\mathrm{s}}=-a\left(dI_{\mathrm{b}}/dz\right)$
that represents an equivalent distributed current generator \cite{marcuvitz1951representation,tamma2014extension,rouhi2021exceptional,rouhi2023modeling}.
This current generator accounts for the effect of the beam's charge
wave flowing in the SWS. It is well known that dependent sources are
used to describe gain in transistors and linear amplifiers, which
justifies this approach to model the e-beams effect on the TL. The
frequency dependent parameters $Z$ and $Y$ could be obtained using
the cascaded circuit model described in Subsection \ref{subsec:Cascaded-Circuit-Model}
as follows. We evaluate the phase velocity of the cold circuit EM
modes $v_{\mathrm{c}}\left(\omega\right)=\omega/\beta_{\mathrm{c}}\left(\omega\right)$,
where $\beta_{\mathrm{c}}\left(\omega\right)=\sqrt{-Z\left(\omega\right)Y\left(\omega\right)}$
is the phase propagation constant harmonic of the cold SWS mode interacting
with the e-beam, and the equivalent TL characteristic impedance $Z_{\mathrm{c}}\left(\omega\right)$.
Then, one could obtain the equivalent frequency-dependent distributed
series impedance $Z\left(\omega\right)=j\beta_{\mathrm{c}}\left(\omega\right)Z_{\mathrm{c}}\left(\omega\right)$
and shunt admittance $Y\left(\omega\right)=j\beta_{\mathrm{c}}\left(\omega\right)/Z_{\mathrm{c}}\left(\omega\right)$.

For convenience, we define a state vector $\boldsymbol{\Psi}\left(z\right)=\left[V,I,V_{\mathrm{b}},I_{\mathrm{b}}\right]^{\mathrm{T}}$ that describes the hot
mode propagation, and rewrite (\ref{eq: V_b}), (\ref{eq: I_b}),
(\ref{eq: TelegV}), and (\ref{eq: TelegI}) in matrix form as

\begin{equation}
\frac{d}{dz}\boldsymbol{\Psi}(z)=-j\mathbf{\underline{M}}\boldsymbol{\Psi}(z),\label{eq: System Equation}
\end{equation}

\begin{equation}
\underline{\mathbf{M}}=\left[\begin{array}{rrrr}
0 & \beta_{\mathrm{c}}Z_{\mathrm{c}} & 0 & 0\\
\beta_{\mathrm{c}}/Z_{\mathrm{c}} & 0 & -ag & -a\beta_{0}\\
0 & a\beta_{\mathrm{c}}Z_{\mathrm{c}} & \beta_{0} & \zeta_{\mathrm{sc}}\\
0 & 0 & g & \beta_{0}
\end{array}\right],\label{eq: System Matrix}
\end{equation}
where $\underline{\mathbf{M}}$ is the $4\times4$ system matrix.
Here, we have used directly the primary TL parameters $\beta_{\mathrm{c}}\left(\omega\right)$
and $Z_{\mathrm{c}}\left(\omega\right)$ instead of $Z\left(\omega\right)$
and $Y\left(\omega\right)$. In the above system matrix, $\zeta_{\mathrm{sc}}$
is the space-charge parameter related to the debunching of beam's
charges, and is given by \cite{rouhi2021exceptional}

\begin{equation}
\zeta_{\mathrm{sc}}=\frac{R_{\mathrm{sc}}}{A\varepsilon_{0}\omega}=\frac{2V_{0}\omega{}_{\mathrm{q}}^{2}}{\omega I_{0}u_{0}},
\end{equation}
where $\omega_{\mathrm{q}}=R_{\mathrm{sc}}\omega_{\mathrm{p}}$ is
the reduced plasma angular frequency, $\omega_{\mathrm{p}}=\sqrt{-\rho_{0}\eta/\left(A\varepsilon_{0}\right)}=\sqrt{I_{0}u_{0}/\left(2V_{0}A\varepsilon_{0}\right)}$
is the plasma frequency \cite{hammer1967coupling}, and $R_{\mathrm{sc}}$
is the plasma frequency reduction factor \cite{branch1955plasma,booske2004insights}.
The term $R_{\mathrm{sc}}$ accounts for reductions in the magnitude
of the axial component of the space-charge electric field due to either
a finite beam radius or proximity to the surrounding conducting walls
of the e-beam tunnel \cite{antonsen1998traveling} (details in Appendix
\ref{app:Plasma-Reduction-Factor}). As shown in Appendix \ref{app:Coupling-Strength-Coefficient},
the coupling strength coefficient $a$ is found by using the formula

\noindent 
\begin{equation}
a=\sqrt{\frac{Z_{\mathrm{P,e}}}{Z_{\mathrm{c}}}},\label{eq:a-MainDef}
\end{equation}
and it is frequency dependent as shown later on. In summary, all the
parameters of the presented model are found using cold simulations
of the EM mode in the serpentine waveguide SWS to estimate the performance
of the hot structure. We emphasize that the calculated characteristic
impedance $Z_{\mathrm{c}}$, regardless of how it is defined, yields
meaningful results in our theoretical model, as long as one uses the
effective interaction impedance $Z_{\mathrm{P,e}}$ that is calculated
from full-wave eigenmode simulations as described in Section \ref{sec:Pierce-Impedance}.

\subsection{Characteristic Equation and Electronic Beam Admittance}

Assuming a state vector $z$-dependence of the form $\boldsymbol{\Psi}\left(z\right)\propto\exp\left(-jkz\right)$,
where $k$ is the complex-valued wavenumber of a hot mode in the interactive
system, leads to the eigenvalue problem $k\boldsymbol{\Psi}\left(z\right)=\mathbf{\underline{M}}\boldsymbol{\Psi}\left(z\right)$.
The resulting modal dispersion characteristic equation is given by

\begin{equation}
\begin{array}{c}
D\left(\omega,k\right)=\det\left(\mathbf{\underline{M}}-k\mathbf{\underline{I}}\right)=k^{4}-k^{3}\left(2\beta_{0}\right)\\
\\
+k^{2}\left(\beta_{0}^{2}-\beta_{\mathrm{q}}^{2}-\beta_{\mathrm{c}}^{2}+a^{2}g\beta_{\mathrm{c}}Z_{\mathrm{c}}\right)+k\left(2\beta_{\mathrm{c}}^{2}\beta_{0}\right)\\
\\
-\beta_{\mathrm{c}}^{2}\left(\beta_{0}^{2}-\beta_{\mathrm{q}}^{2}\right)=0,
\end{array}\label{eq:a-main-def}
\end{equation}
where $\beta_{\mathrm{q}}=\omega_{\mathrm{q}}/u_{0}=\sqrt{g\zeta_{\mathrm{sc}}}$
is the phase constant of space-charge wave. The solution of (\ref{eq:a-main-def})
leads to four modal complex-valued wavenumbers of the four hot modes
in the interactive system. The characteristic equation is rewritten
as follows
\begin{figure}[t]
\begin{centering}
\includegraphics[width=0.85\columnwidth]{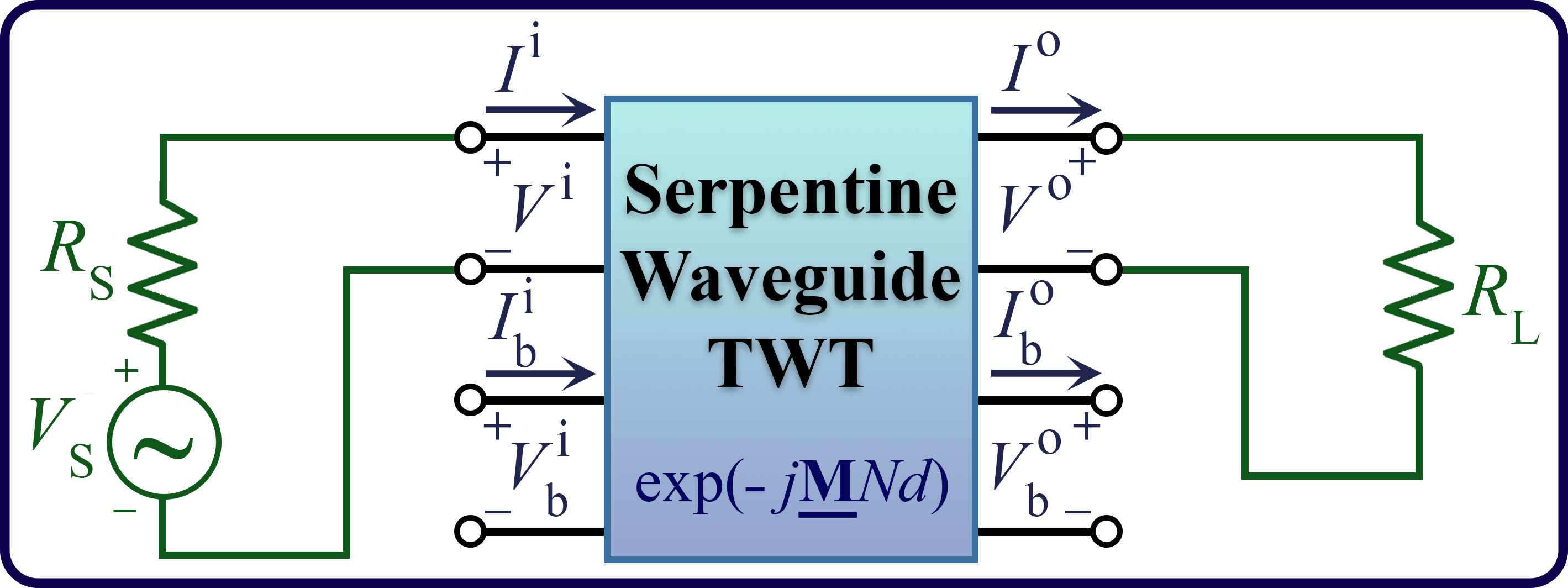}
\par\end{centering}
\centering{}\caption{Circuit model for gain calculation considering frequency-dependent
resistances for the source and load ($R_{\mathrm{S}}$ and $R_{\mathrm{L}}$).\label{fig: GainCalculation}}
\end{figure}

\begin{equation}
\left(k^{2}-\beta_{\mathrm{c}}^{2}\right)\left[\left(k-\beta_{0}\right)^{2}-\beta_{\mathrm{q}}^{2}\right]=-a^{2}gk^{2}\beta_{\mathrm{c}}Z_{\mathrm{c}},\label{eq: Dispersion2}
\end{equation}
to stress that the term $-a^{2}g\beta_{\mathrm{c}}Z_{\mathrm{c}}k^{2}$
($=-g\beta_{\mathrm{c}}Z_{\mathrm{P,e}}k^{2}$) indicates the coupling
between the two dispersion equations of the isolated EM waves in the
cold SWS $\left(k^{2}-\beta_{\mathrm{c}}^{2}\right)=0$, and isolated
charge waves $\left[\left(k-\beta_{0}\right)^{2}-\beta_{\mathrm{q}}^{2}\right]=0$.
Here, only parameters obtained from cold SWS simulations are used
to find the dispersion of the four hot modes. For a given eigenmode,
the e-beam interaction with the EM wave could be completely modeled
as an active TL with a voltage-dependent current source, as shown
schematically in Fig. \ref{fig: TLHot}, given by \cite{tamma2014extension}

\begin{equation}
i_{\mathrm{s}}=jakI_{\mathrm{b}}=-Y_{\mathrm{b}}V,
\end{equation}
where the electronic beam admittance per unit length $Y_{\mathrm{b}}$
is

\begin{equation}
Y_{\mathrm{b}}=-j\frac{a^{2}gk^{2}}{\left[\left(k-\beta_{0}\right)^{2}-\beta_{\mathrm{q}}^{2}\right]}.\label{eq:EectrBeamAdmit}
\end{equation}
This admittance is a generalization of the one already provided in
\cite{tamma2014extension} since here we have included the space charge
effect $\beta_{\mathrm{q}}^{2}=g\zeta_{\mathrm{sc}}$.

\begin{figure*}[tbh]
\begin{centering}
\includegraphics[width=1\textwidth]{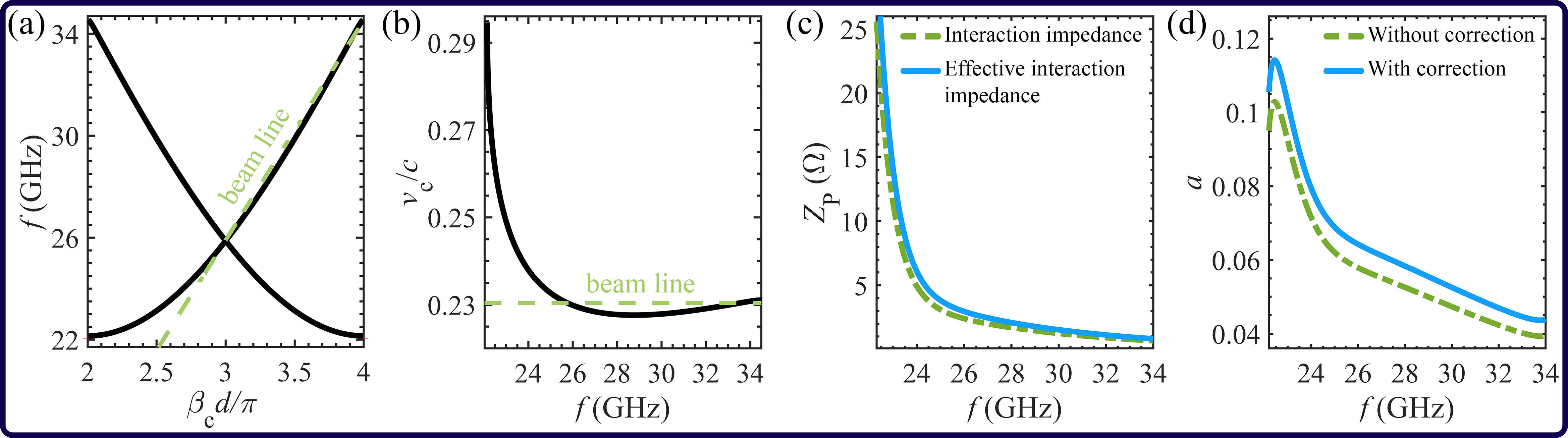}
\par\end{centering}
\centering{}\caption{Cold results: (a) Dispersion curve and (b) normalized phase velocity
of the modes in the cold serpentine waveguide SWS in the first spatial
harmonic. (c) The on-axis interaction impedance of the serpentine
waveguide SWS at the center of the beam tunnel with (solid blue) and
without (dashed green) considering the correction factor $\delta_{\mathrm{e}}$.
(d) The frequency-dependent value of the coupling strength coefficient
$a$ with (solid blue) and without (dashed green) correction factor
$\delta_{\mathrm{e}}$.\label{fig:ColdforHotExample}}
\end{figure*}

\subsection{TWT Amplifier Gain \label{subsec:Gain-Versus-Frequency}}

We describe the theoretical calculation to compute the gain of a TWT
amplifier using the circuit model illustrated in Fig. \ref{fig: GainCalculation},
where a matched resistance is considered for the source generator
$R_{\mathrm{S}}$, and the output is terminated by the matched load
$R_{\mathrm{L}}$. The serpentine waveguide TWT is modeled by the
system matrix $\underline{\mathbf{M}}$ described earlier, input state
vector of $\boldsymbol{\Psi}_{1}=\left[V^{\mathrm{i}},I^{\mathrm{i}},V_{\mathrm{b}}^{\mathrm{i}},I_{\mathrm{b}}^{\mathrm{i}}\right]^{\mathrm{T}}$
calculated at $z=0$, and output state vector of $\boldsymbol{\Psi}_{2}=\left[V^{\mathrm{o}},I^{\mathrm{o}},V_{\mathrm{b}}^{\mathrm{o}},I_{\mathrm{b}}^{\mathrm{o}}\right]^{\mathrm{T}}$
calculated at $z=Nd$, i.e., at the end of the TWT, where $N$ indicates
the number of unit cells. The output state vector is calculated as
$\boldsymbol{\Psi}_{2}=\underline{\mathbf{T}}\boldsymbol{\Psi}_{1}$,
where $\underline{\mathbf{T}}=\exp\left(-j\underline{\mathbf{M}}Nd\right)$
is the TWT transfer matrix.

In the model, we use the following boundary conditions at $z=0$ and
$z=Nd$,

\begin{equation}
\left\{ \begin{array}{l}
V_{\mathrm{b}}^{\mathrm{i}}=0,\;I_{\mathrm{b}}^{\mathrm{i}}=0\\
V^{\mathrm{i}}+I^{\mathrm{i}}R_{\mathrm{S}}=V_{\mathrm{S}},\;V^{\mathrm{o}}-I^{\mathrm{o}}R_{\mathrm{L}}=0
\end{array}\right.\label{eq: BC1}
\end{equation}
In these equations, the source resistance $R_{\mathrm{S}}$ and load
resistance $R_{\mathrm{L}}$ are assumed to be equal to the frequency-dependent
characteristic impedance of the serpentine waveguide $Z_{\mathrm{c}}$,
and $V_{\mathrm{S}}$ is the generator voltage source. We solve the
system of equations at each frequency and calculate the equivalent
circuit current and voltage (proportional to the electric and magnetic
fields) at the TWT output port. Then, we calculate the output power
$P_{\mathrm{out}}=\left|V^{\mathrm{o}}\right|^{2}/\left(2R_{\mathrm{L}}\right)$,
and the available input power $P_{\mathrm{av}}=\left|V_{\mathrm{S}}\right|^{2}/\left(8R_{\mathrm{S}}\right)$
(also denoted as incident power) to obtain the frequency-dependent
gain as $G=P_{\mathrm{out}}/P_{\mathrm{av}}$.

In order to calculate the gain, we build the linear system $\mathbf{\underline{A}}\mathbf{X}=\mathbf{B}$,
where the vector $\mathbf{X}=\left[V^{\mathrm{i}},I^{\mathrm{i}},V_{\mathrm{b}}^{\mathrm{i}},I_{\mathrm{b}}^{i},V^{\mathrm{o}},I^{\mathrm{o}},V_{\mathrm{b}}^{\mathrm{o}},I_{\mathrm{b}}^{\mathrm{o}}\right]^{\mathrm{T}}$
contains the state vectors at the input and output of the TWT, and
the $8\times8$ matrix $\mathbf{\underline{A}}$ is defined as

\noindent 
\begin{equation}
\underline{\boldsymbol{\mathrm{A}}}=\left[\begin{array}{cc}
\left[-\exp\left(-j\underline{\mathbf{M}}Nd\right)\right] & \left[\underline{\boldsymbol{\mathrm{I}}}_{4}\right]\\
\begin{array}{cccc}
0 & 0 & 1 & 0\\
0 & 0 & 0 & 1\\
1 & R_{\mathrm{S}} & 0 & 0\\
0 & 0 & 0 & 0
\end{array} & \begin{array}{cccc}
0 & 0 & 0 & 0\\
0 & 0 & 0 & 0\\
0 & 0 & 0 & 0\\
1 & -R_{\mathrm{L}} & 0 & 0
\end{array}
\end{array}\right],\label{eq:Gain_A}
\end{equation}

\noindent where $\underline{\boldsymbol{\mathrm{I}}}_{4}$ is the
$4\times4$ identity matrix. The input vector of the system is expressed
as $\mathbf{B}=\left[0,0,0,0,0,0,V_{\mathrm{S}},0\right]^{\mathrm{T}}$.
Then, solving this $8\times8$ system of equations for the vector
$\mathbf{X}$ allows us to compute the TWT gain.

\section{Validation of Model for Hot Structure\label{sec:Validation-of-Hot}}

In order to investigate the accuracy of the presented model for the
interaction, we compare the theoretically calculated gain versus frequency
results from our model with those numerically obtained from the commercial
PIC software CST Particle Studio. As explained in Section \ref{sec:Pierce-Impedance},
in our model we consider the effective interaction impedance, which
describes the strength of beam-EM mode interaction in the TWT. In
this paper, the correction factor $\delta_{\mathrm{e}}$ is calculated
by matching the maximum gain value from the theoretical model with
the one obtained by \textit{only one} PIC simulation that occurs at
the synchronization frequency, $Z_{\mathrm{P}}$ is determined from
(\ref{eq:Zpierce}) by post-processing the data extracted from CST
eigenmode simulations and $v_{\mathrm{c}}$ is calculated by theoretical
circuit models, i.e., a cascaded circuit model.

In this study, synchronization with the first spatial harmonic of
the SWS is selected for low beam voltage operation. However, for simplicity
of notation, we drop the harmonic index number and we will call the
circuit modal wavenumber and phase velocity belonging to the $m=1$
spatial harmonic simply as $\beta_{\mathrm{c}}$ and $v_{\mathrm{c}}$.
We consider a serpentine waveguide SWS with the geometry parameters
listed in Table \ref{tab:Parameters}. The e-beam has $I_{0}=10\:\mathrm{mA}$
and a radius $r_{\mathrm{b}}=0.35\:\mathrm{mm}$ and we end up with
a tunnel filling factor of $\left(r_{\mathrm{b}}/r_{\mathrm{c}}\right)^{2}=0.5$.
For the e-beam, the normalized phase velocity $u_{0}/c$ is set to
be $0.230$. This value corresponds to an average kinetic voltage
of $V_{0}=14.077\:\mathrm{kV}$ for the e-beam. Additionally, a uniform
longitudinal magnetic field of $0.8\;\mathrm{T}$ was applied to confine
the e-beam. The cold dispersion diagram and beam line are illustrated
in Fig. \ref{fig:ColdforHotExample}(a) where the beam line with normalized
phase velocity of $u_{0}/c=0.230$ is superimposed to the wavenumber
of the EM mode, in the SWS on both left and right of the $3\pi$ point.
Additionally, the beam line may synchronize with the EM backward mode
near $3\pi$ at the intersection frequency which may result in parasitic
oscillations and instability \cite{nguyen2014design,hung2015absolute}.
So, in the design of a long serpentine waveguide TWT, attenuators
can be used to mitigate oscillation risk. However, this issue is not
discussed here, and how the presented model can be adapted to cases
with attenuators will be studied in our future work. The frequency-dependent
interaction impedance calculated at the beam center for the first
spatial harmonic by using (\ref{eq:Zpierce}) is shown in Fig. \ref{fig:ColdforHotExample}(c).
In this example, the interaction impedance correction factor is considered
to be $\delta_{\mathrm{e}}=0.11$, which is a relatively small factor.
We show both the calculated interaction impedance without correction
factor $Z_{\mathrm{P}}$ (see (\ref{eq:Zpierce})) and effective interaction
impedance with correction factor $Z_{\mathrm{P,e}}$ (see (\ref{eq:ZPierceCorrectDelta}))
in Fig. \ref{fig:ColdforHotExample}(c) by using dashed green and
solid blue curves respectively. The on-axis interaction impedance
approaches very high values near the waveguide cutoff frequency at
$f_{\mathrm{c}}=22.15\:\mathrm{GHz}$ and gradually drops as the frequency
grows further away from the cutoff frequency. The frequency-dependent
value of the coupling strength coefficient without considering correction
factor, $a=\sqrt{Z_{\mathrm{P}}/Z_{\mathrm{c}}}$, and with correction
factor, $a=\sqrt{Z_{\mathrm{P,e}}/Z_{\mathrm{c}}}=(1+\delta_{\mathrm{e}})\sqrt{Z_{\mathrm{P}}/Z_{\mathrm{c}}}$,
are shown in Fig. \ref{fig:ColdforHotExample}(d).
\begin{figure}[t]
\begin{centering}
\includegraphics[width=1\columnwidth]{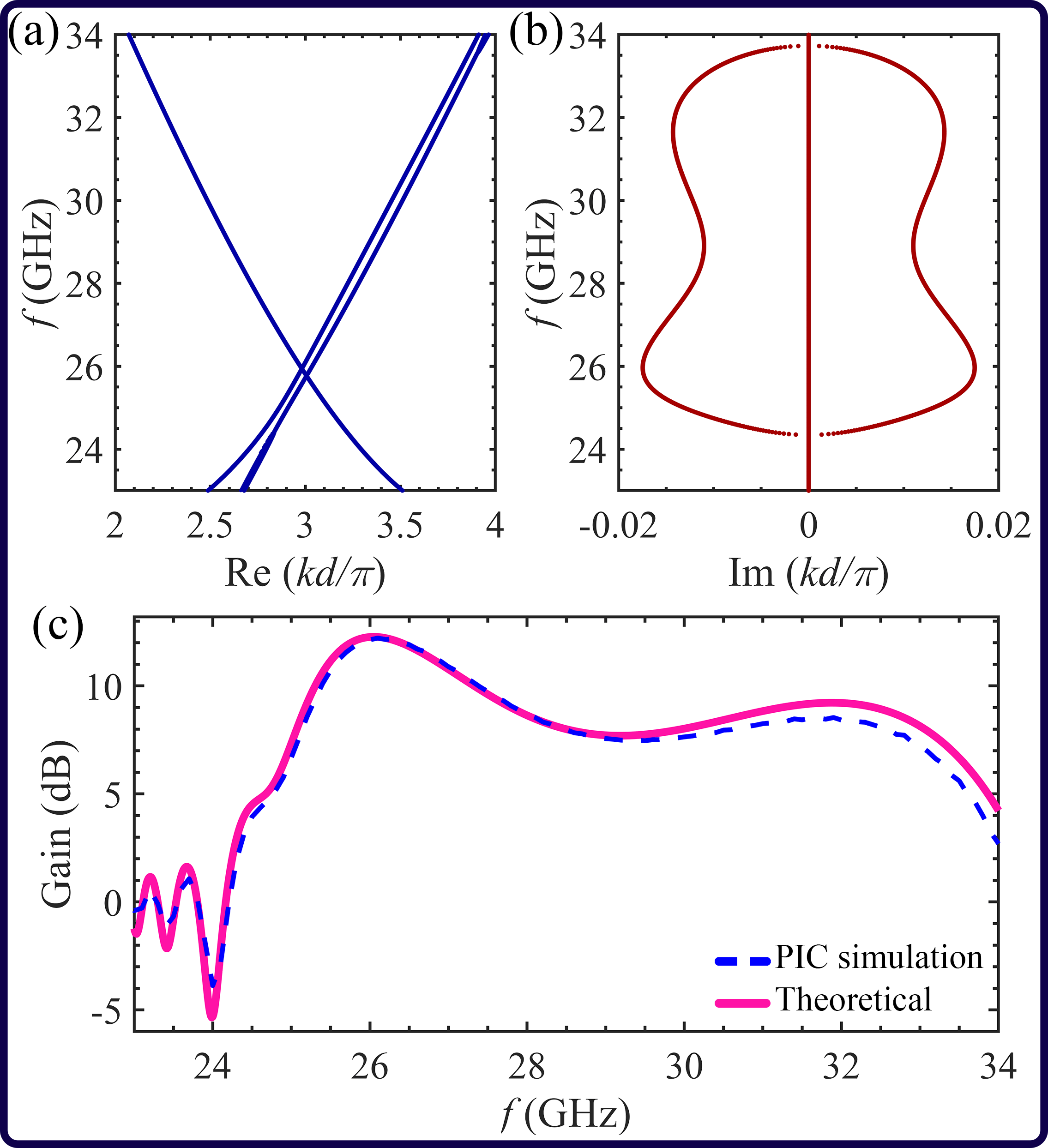}
\par\end{centering}
\centering{}\caption{Hot results: The (a) real and (b) imaginary parts of complex-valued
wavenumbers of hot modes by varying frequency. (c) TWT gain versus
frequency predicted by the proposed theoretical model (solid pink),
compared to 3D PIC simulations (dashed blue).\label{fig:HotforHotExample}}
\end{figure}

According to the intersection of the cold EM mode phase velocity curve
$v_{\mathrm{c}}$ and the beam line in Fig. \ref{fig:ColdforHotExample}(b),
we observe beam-EM wave full synchronization at $25.73\:\mathrm{GHz}$
and $33.52\:\mathrm{GHz}$, where high amplification is expected to
occur. The real and imaginary parts of the complex-valued wavenumber
of the hot modes (i.e., accounting for the beam-EM wave interaction)
are calculated by (\ref{eq: Dispersion2}) and shown in Figs. \ref{fig:HotforHotExample}(a)
and (b). The amplification regime is obtained when there is a hot
mode with $\mathrm{Im}\left(k\right)>0$. The numerical gain versus
frequency diagram is theoretically calculated by the method described
in Subsection \ref{subsec:Gain-Versus-Frequency} for the serpentine
waveguide TWT with $N=40$ unit cells ($160\:\mathrm{mm}$ in length)
and input power of $P_{\mathrm{in}}=0\:\mathrm{dBm}$. It is compared
with the one obtained by computationally intensive 3D PIC simulations,
resulting in excellent agreement. The comparison also validated the
value of the interaction impedance correction factor $\delta_{\mathrm{e}}=0.11$.
Since the analysis is in the linear regime, instead of using $N=40$
unit cells, a quick simulation to estimate the correction factor $\delta_{\mathrm{e}}$
was done based on only $N=10$ unit cells. However, as a check we
also verified that we obtained the same value for correction factor when considering $N=40$ unit cells.

The theoretical and PIC simulated gain versus frequency are illustrated
in Fig. \ref{fig:HotforHotExample}(c) by solid pink and dashed blue
curves, respectively. The agreement is excellent, indicating the accuracy
of the model. Additionally, as predicted, maximum gains are obtained
around synchronization frequencies. The total number of mesh cells
in the simulation is approximately 2.6 million and a steady state
output signal is seen after a transient time of $10\:\mathrm{ns}$
elapses. We use a sinusoidal signal as an excitation signal in the
PIC simulation and a frequency sweep is performed to calculate output
power in the selected frequency band. The required time for simulation
and specification of the employed server is provided in Appendix \ref{sec:Computational-Burden-and}.
As shown in Fig. \ref{fig:HotforHotExample}(c), the 3-dB bandwidth
is $9.37\%$ covering from $25.21\:\mathrm{GHz}$ to $27.65\:\mathrm{GHz}$.
Also, the maximum amplifier gain of $12.27\:\mathrm{dB}$ is obtained
at $26.04\:\mathrm{GHz}$. We also investigated another example with
a wider e-beam with tunnel filling factor of $\left(r_{\mathrm{b}}/r_{\mathrm{c}}\right)^{2}=0.95$.
In this case, the correction factor is $\delta_{\mathrm{e}}=0.18$.
This value is explainable since according to Fig. \ref{fig: NonunimformElectricField}(c)
and (d) we observe bigger values of electric fields near the beam
tunnel wall which leads to stronger beam-EM wave interaction. Note
that the purpose of this paper is not to design a TWT that can compete
with conventional designs, but to showcase a simple and accurate model
to predict TWT performance.

\section{Parameter Study\label{sec:Parameter-Study}}

\begin{figure}[t]
\begin{centering}
\includegraphics[width=1\columnwidth]{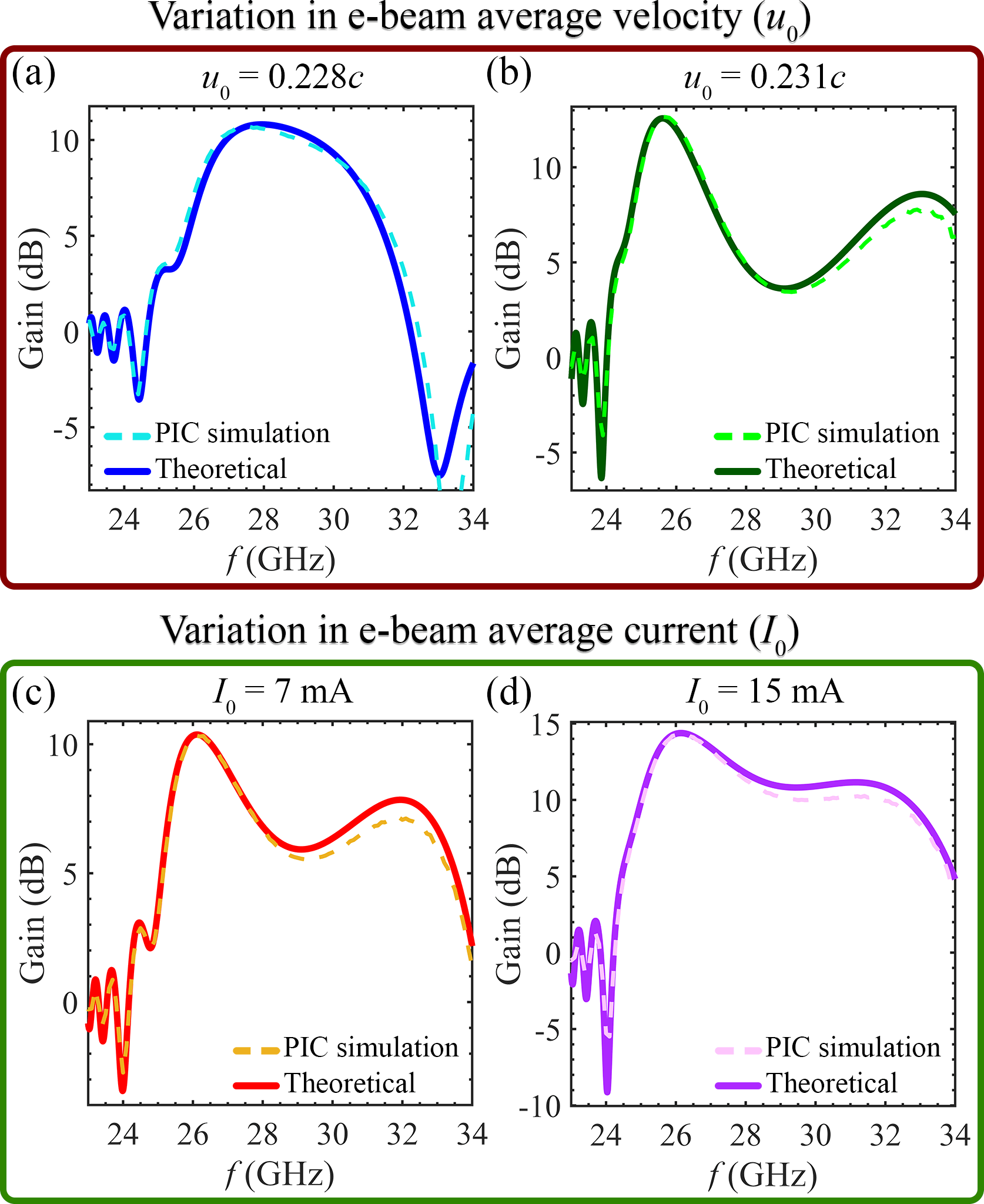}
\par\end{centering}
\centering{}\caption{Comparison of gain versus frequency for a serpentine waveguide TWT
calculated using our theoretical model and PIC simulations. In the
first row, we show gain diagram by varying e-beam average phase velocity
as (a) $u_{0}=0.228c$ and (b) $u_{0}=0.231c$. In the second row,
we illustrate gain diagram by varying e-beam average current as (c)
$I_{0}=7\:\mathrm{mA}$ and (d) $I_{0}=15\:\mathrm{mA}$. The dashed
curves in these plots show the results obtained via PIC simulation
whereas solid curves are obtained based on the proposed theoretical
model.\label{fig: ParameterStudy1}}
\end{figure}

To validate the presented model, a variety of simulation runs and
comparisons have been carried out. We will apply the same correction
factor $\delta_{\mathrm{e}}=0.11$ obtained in the previous section
to all the following examples. In fact, the effective interaction
impedance and correction factor $\delta_{\mathrm{e}}$ are identical
for all examples, even when changing the e-beam parameters, number
of unit cells and input power in the linear regime. First, we vary
$u_{0}$ to change the synchronization frequency but leave all other
parameters unchanged, which are equal to the parameters used in Section
\ref{sec:Validation-of-Hot}. In Fig. \ref{fig: ParameterStudy1}(a),
we select $u_{0}=0.228c$, which is $0.002c$ slower than the value
used in the previous example. In this case, the forward branch of
the modal dispersion diagram is approximately linear in the vicinity
of the optimum frequency (i.e., the phase velocity remains almost
constant). Here, the 3-dB bandwidth is $15.87\%$ of the center frequency
covering from $26.27\:\mathrm{GHz}$ to $30.70\:\mathrm{GHz}$ and
the maximum amplifier gain of $10.82\:\mathrm{dB}$ is predicted at
$27.93\:\mathrm{GHz}$. Consequently, by establishing optimum synchronization,
we can dramatically increase bandwidth. Next, in Fig. \ref{fig: ParameterStudy1}(b)
we increase the e-beam phase velocity to $u_{0}=0.231c$, which leads
to synchronization around $f_{\mathrm{sync}}=25.5\:\mathrm{GHz}$
and $f_{\mathrm{sync}}=34.26\:\mathrm{GHz}$, and calculate the gain.
In these two plots, we also illustrate the theoretically calculated
gain based on the proposed theoretical method, and we observe excellent
agreement between theoretical (solid curves) and PIC simulation (dashed
curves) results. We stress that we did not have to recalculate the
correction factor $\delta_{\mathrm{e}}$ that was already calculated
in the example in the previous section.

\begin{figure}[t]
\begin{centering}
\includegraphics[width=1\columnwidth]{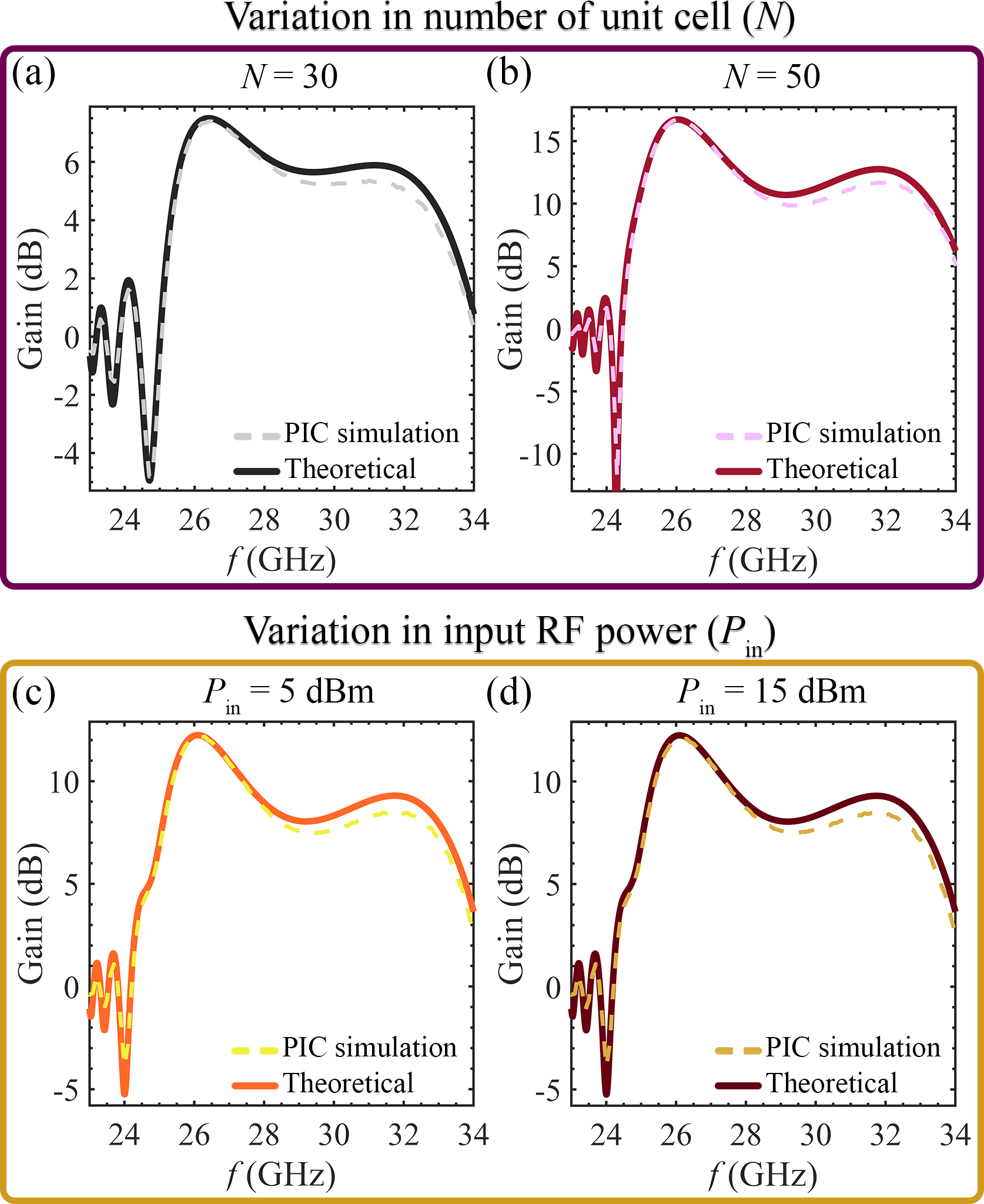}
\par\end{centering}
\centering{}\caption{Comparison of gain versus frequency for a serpentine waveguide TWT
calculated using our theoretical model and PIC simulations. In the
first row, we show gain diagram by varying number of unit cells as
(a) $N=30$ and (b) $N=50$. In the second row, we illustrate gain
diagram by varying input power as (c) $P_{\mathrm{in}}=5\:\mathrm{dBm}$
and (d) $P_{\mathrm{in}}=15\:\mathrm{dBm}$. The dasehd curves in these
plots show the results obtained via PIC simulation whereas solid
curves are obtained based on the proposed theoretical model.\label{fig: ParameterStudy2}}
\end{figure}

In the next step, the gain diagrams are calculated for the e-beam
average currents of $I_{0}=7\:\mathrm{mA}$ and $I_{0}=15\:\mathrm{mA}$,
shown in Figs. \ref{fig: ParameterStudy1}(c) and (d). All the other
parameters are as described in the previous section. The maximum gain
in both cases occurs approximately at the same frequency since the
e-beam phase velocity is equal in both examples. On the other hand,
the maximum gain for the current value of $I_{0}=15\:\mathrm{mA}$
is much bigger than the gain value for $I_{0}=7\:\mathrm{mA}$. Hence,
it is critical to choose the proper value for the e-beam current to
avoid saturation. The solid curves obtained based on the proposed
theoretical model show good agreement with the dashed curves calculated
using PIC simulation. It is important to note that the correction
factor $\delta_{\mathrm{e}}=0.11$ that was calculated in the previous
section did not need to be adjusted or recalculated.

Our next step is to demonstrate how selecting the number of unit cells
affects the gain and how this gain can be calculated accurately by
the proposed model, still retaining the same correction factor $\delta_{\mathrm{e}}=0.11$
that was already calculated in the example in the previous section.
The gain diagrams by varying the number of unit cells as $N=30$ and
$N=50$ are calculated and shown in Figs. \ref{fig: ParameterStudy2}(a)
and (b). In both cases, the e-beam has the same phase velocity, so
maximum gain occurs roughly at the same frequency. The solid curves
calculated by the theoretical model show excellent agreement with
the dashed curves obtained by numerically intensive PIC simulations.
Increasing the number of interaction unit cells too much will eventually
result in undesirable oscillations when the small-signal gain becomes
too high (e.g. above the practical limit of 30 dB for a single-stage
TWT). Therefore, it is critical to consider the proper number of unit
cells to prevent oscillations. As a result of using the longer device
for higher gain extraction, we should use sever in the design which
will be investigated in detail in our future work.

Next, we show the effect of input power variation on the gain diagram,
but still retaining the same correction factor $\delta_{\mathrm{e}}=0.11$
that was already calculated in the example in the previous section.
Since our method is based on small-signal approximation, we neglected
the effect of nonlinear terms in our model. We provide two different
examples with $P_{\mathrm{in}}=5\:\mathrm{dBm}$ and $P_{\mathrm{in}}=15\:\mathrm{dBm}$
and the calculated results are presented in Figs. \ref{fig: ParameterStudy2}(c)
and (d). In comparison to the dashed curves obtained by PIC simulations,
the theoretical results represented by solid curves exhibit good agreement.

\begin{figure*}[tbh]
\begin{centering}
\includegraphics[width=0.65\textwidth]{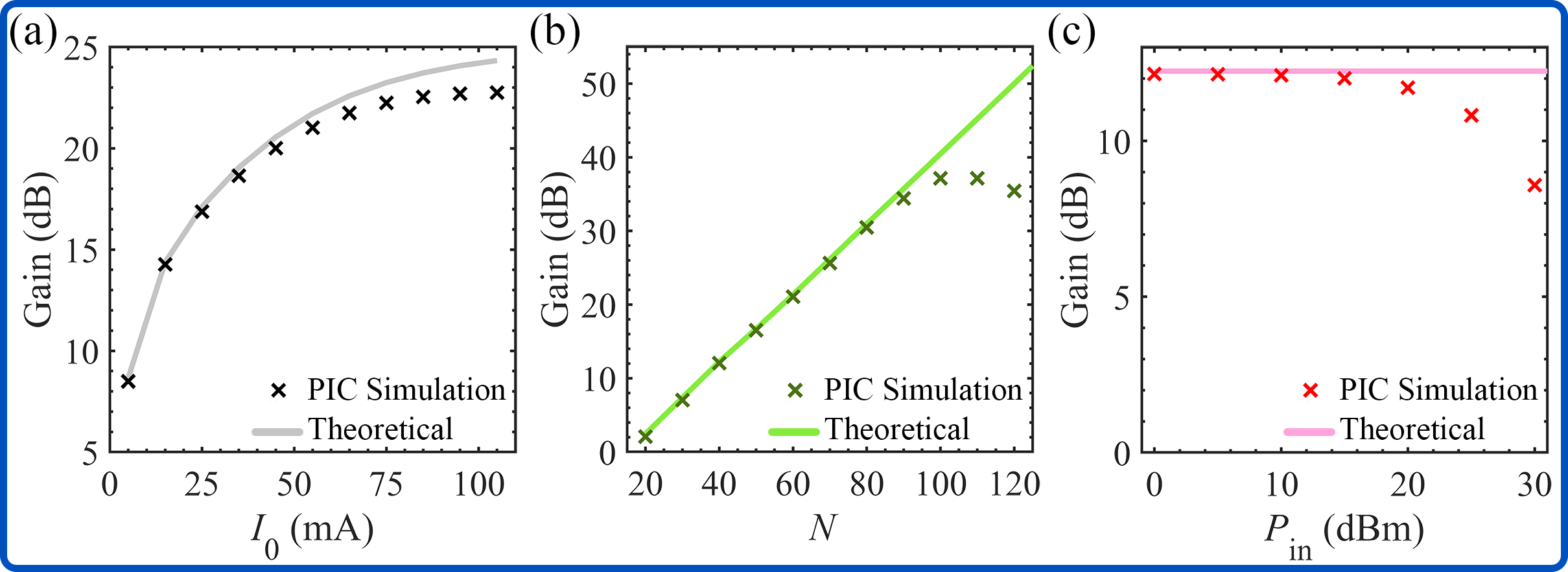}
\par\end{centering}
\centering{}\caption{Comparison of gain versus frequency for a serpentine waveguide TWT
calculated using our developed theoretical model (solid curves) and
PIC simulations (dashed curves). In these plots, we change (a) e-beam
average current, (b) number of unit cells, and (c) input power to
show the accuracy of our theoretical calculation in the linear regime.\label{fig: ParameterStudy3}}
\end{figure*}

As a last analysis, we calculate the gain at the synchronization frequency
of $f_{\mathrm{sync}}=26\:\mathrm{GHz}$ by assuming the parameters
used in Section \ref{sec:Validation-of-Hot}. The gain diagram by
varying the e-beam average current is shown in Fig. \ref{fig: ParameterStudy3}(a).
In this plot, the theoretical gain is shown by a solid curve and the
cross sign shows the corresponding simulation gain obtained from PIC
simulation at sampled currents. When the e-beam current is increased,
saturation occurs, so designers should choose the proper current value
carefully. The analogous analysis is provided by varying the number
of unit cells and the calculated gain at the synchronization frequency
by both theoretical and PIC simulation is shown in Fig. \ref{fig: ParameterStudy3}(b).
The simulation results confirm the calculated gain value when the
number of unit cells is lower than 90 elements. It is significant
to note that we did not use sever in the design of TWTs and all the
simulations are provided for single-stage TWT. Finally, Fig. \ref{fig: ParameterStudy3}(c)
shows the linear and saturation regions of the TWT by varying the
incident RF power at the TWT input port (at the cathode end). The
analysis shows that by increasing the input power, we move into a
nonlinear regime, where the calculated gain by using the theoretical
model and PIC simulation disagree by more than 3 dB. In the large
signal regime, the calculated results based on linear approximation
will not reliably reproduce the TWT behavior near saturation. Our
developed model is reliable for the small-signal regime and accurate
results in the large-signal regime should be calculated using other
specialized large-signal codes or PIC simulations. Again, it should
be noted that the correction factor value in these plots has not changed
from previous examples.

\begin{figure}[t]
\begin{centering}
\includegraphics[width=1\columnwidth]{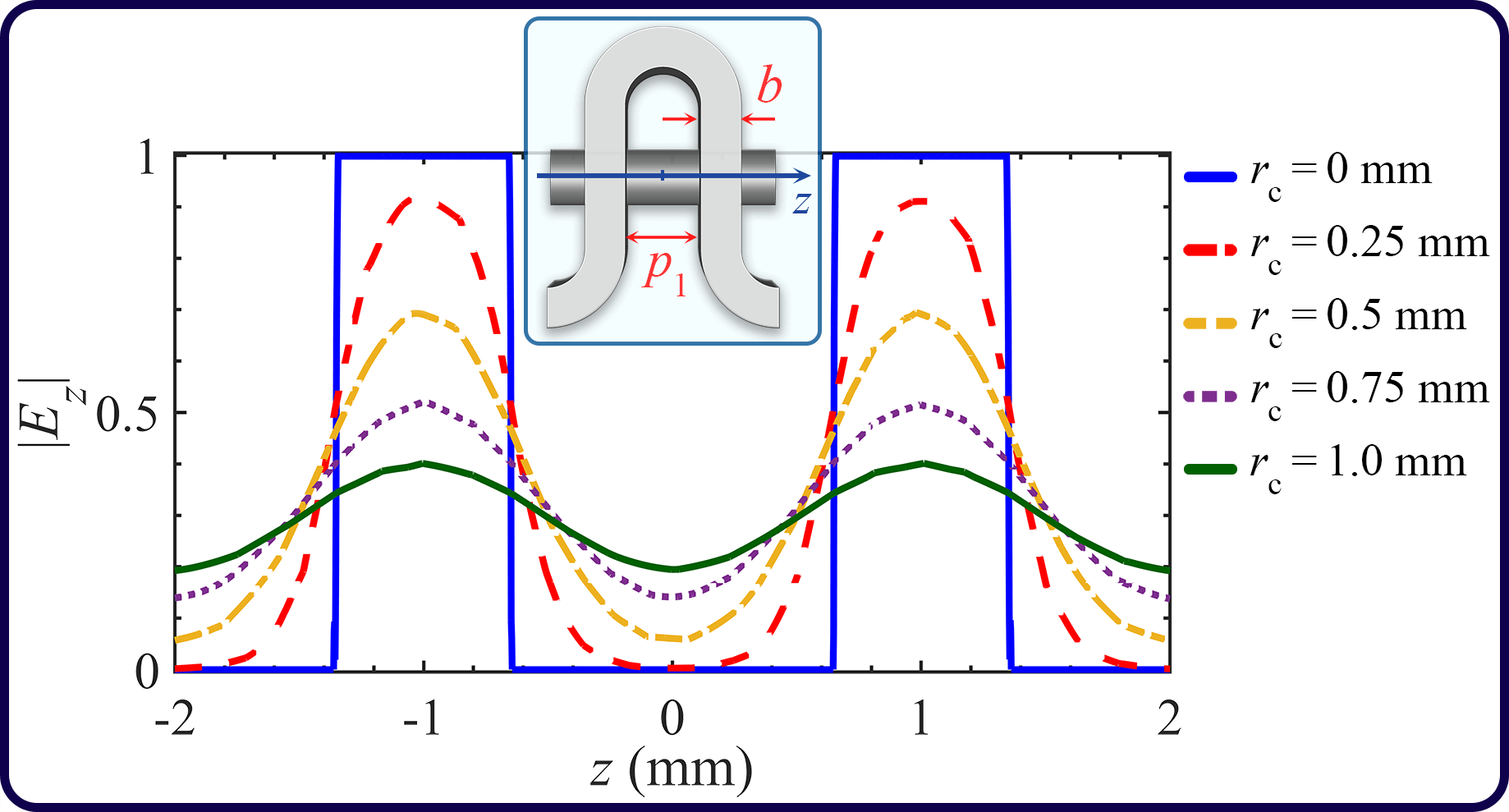}
\par\end{centering}
\centering{}\caption{The on-axis $z$-component of electric field distribution as a function
of longitudinal position $z$ in a one unit cell with $d=4\:\mathrm{mm}$,
at the center of beam tunnel (i.e., $r=0$) for five beam tunnel radii
$r_{\mathrm{c}}$. The calculated values are normalized to the maximum
value of the $z$-component of the electric field in the interaction
area when $r_{\mathrm{c}}=0$, i.e., the case without tunnel.\label{fig: NonunimformElectricField2}}
\end{figure}

\section{Conclusion\label{sec:Conclusion}}

We have presented an extended analytical model for studying beam-EM
wave interaction in a serpentine waveguide TWT that considers space-charge
effects and dispersive waveguide parameters to predict gain in TWT
amplifiers. Our goal is not to present a novel design method but rather
to construct an accurate and robust small-signal model to predict
TWT performance that could also be used for design. The method is
simple because it uses an equivalent circuit model to calculate the
SWS cold (i) modal wavenumber, (ii) characteristic impedance, and
(iii) the interaction impedance, which are all frequency dependent.
We added a frequency-independent correction factor $\delta_{\mathrm{e}}$
to the interaction impedance, to model the nonuniform beam-EM wave
interaction in the overlapping region of the e-beam and SWS longitudinal
electric field. A theoretical method is used to predict the gain versus
frequency and complex-valued wavenumber of the hot modes, and the
results are compared with numerically intensive PIC simulations. The
proposed method has been found always in good agreement with PIC simulations
and much faster and flexible. For example, the flexibility of our
method has been shown by changing the e-beam parameters, number of
unit cells, and input power and by comparing the theoretical gain
results with numerical gain results based on PIC simulations. The
results consistently showed that our model is accurate and efficient
at predicting serpentine waveguide TWT amplification characteristics.

\appendices{}

\section{Longitudinal Fields in The Beam Tunnel\label{app:Non-cross-Section-Interaction}}

A number of works analyzed the effect of variation in the tunnel gap
between the walls of the waveguide and the effect of thin interaction
gap ($b$ in Fig. \ref{fig: NonunimformElectricField2}) \cite{he2010investigation,he2011investigation,hou2012equivalent,guo2012tapered,hou2013novel,wei2014novel}.
In addition, the tunnel between the straight waveguide sections ($p_{1}=d/2-b$
in Fig. \ref{fig: NonunimformElectricField2}) should be long enough
to prevent the guided EM wave from directly coupling between straight
sections via the beam tunnel, which operates below the cutoff \cite{choi1995folded,ha1998theoretical}.
The analysis of the electric field distribution in the beam tunnel
and interaction area of the cold single unit cell is shown in Fig.
\ref{fig: NonunimformElectricField2}. The parameters used in this
example are the same as those listed in Table \ref{tab:Parameters}
and we illustrate the on-axis $z$-component of electric field at
the center of beam tunnel by varying beam tunnel radius $r_{\mathrm{c}}$.
It should be noted that a large tunnel diameter can reduce the effective
longitudinal field at the center of such a tunnel and hence decrease
the gain (for instance, see the green curve in Fig. \ref{fig: NonunimformElectricField2}).
Hence, the beam tunnel radius should be selected carefully.

\begin{figure}[t]
\begin{centering}
\includegraphics[width=1\columnwidth]{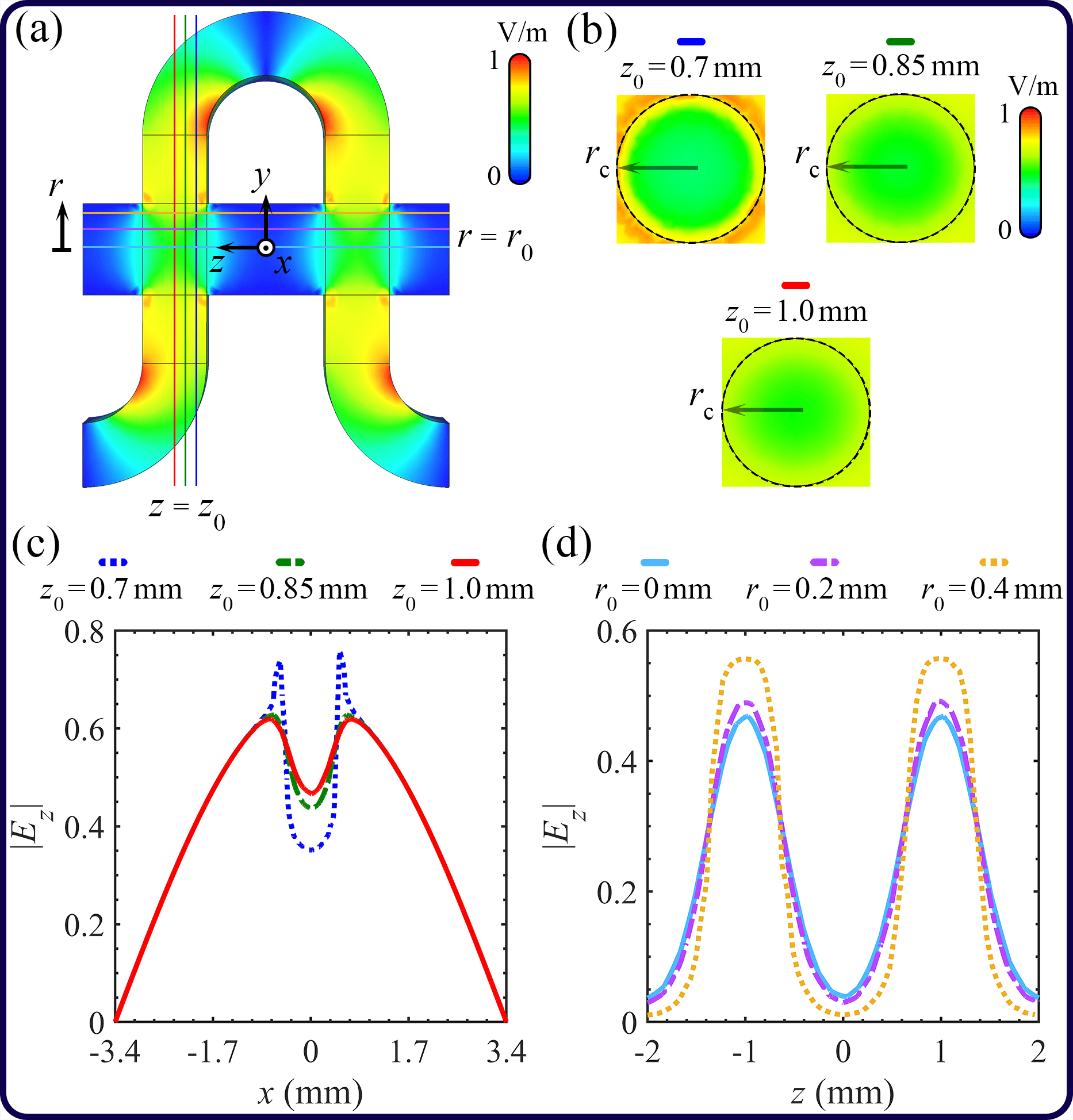}
\par\end{centering}
\centering{}\caption{The on-axis $z$-component of electric field distribution in the beam
tunnel area in a cold serpentine waveguide SWS: (a) Over the longitudinal
cross section, i.e., at $x=0$ plane; (b) in the interaction area
at three different transverse cross section planes with $z_{0}=0.7\:\mathrm{mm}$,
$z_{0}=0.85\:\mathrm{mm}$ and $z_{0}=1\:\mathrm{mm}$; (c) in the
$x$ direction, with $y=0$, at three different $z_{0}$ coordinate;
(d) at three different radii $r_{0}$, along the beam tunnel.\label{fig: NonunimformElectricField}}
\end{figure}

\section{Longitudinal Fields in The Interaction Regions\label{app:Non-uniform-Interaction}}

The magnitude of the $z$-component of electric field distribution
at the center of the longitudinal cross section (the $x=0$ plane)
of a cold serpentine waveguide is shown in Fig. \ref{fig: NonunimformElectricField}(a).
For better illustration, the $z$-component magnitude in the beam
and EM wave interaction area at various transverse cross sections
of $z_{0}=0.7\:\mathrm{mm}$, $z_{0}=0.85\:\mathrm{mm}$ and $z_{0}=1\:\mathrm{mm}$
is shown in Fig.\ref{fig: NonunimformElectricField}(b). We can see
that the electric field magnitude increases near the beam tunnel perimeter.
Also, the magnitude of the $z$-component of the electric field in
a straight line in the $x$ direction, with $y=0$, at three different
$z$ values is depicted in Fig. \ref{fig: NonunimformElectricField}(c).
The magnitude of the $z$-component of the electric field is also
calculated at different radii inside the beam tunnel as shown in Fig.
\ref{fig: NonunimformElectricField}(d). These plots demonstrate that
the minimum field value is obtained at the center of the beam tunnel
and that the magnitude of $E_{z}$ increases gradually with an increasing
radius. Thus, in the interaction area, the minimum interaction impedance
is calculated at the center of the beam tunnel, i.e., at $r=0$. To
account for the nonuniform distribution of the longitudinal electric
field in the interaction area, the interaction impedance should be
multiplied by a correction factor, i.e., $\left(1+\delta_{\mathrm{e}}\right)^{2}$.
Since the electric field magnitude is greater near the tunnel wall,
the correction factor should be greater than one ($\delta_{\mathrm{e}}\geq0$).

\section{Coupling Strength Coefficient\label{app:Coupling-Strength-Coefficient}}

The characteristic impedance of a mode guided by a cold waveguide
is $Z_{\mathrm{c}}$ and by using this value, matching networks can
be designed to terminate the input and output ends of the TWT. In
contrast, in the Pierce model, the characteristic impedance of the
equivalent TL that represents EM synchronization is the interaction
impedance $Z_{\mathrm{P}}$. These two dispersive impedances are related
by a frequency-dependent coupling strength coefficient discussed here.
Other works have used this coupling strength coefficient introduced
as an ad-hoc parameter, including \cite{figotin2013multi,tamma2014extension,Othman2016Theory,Othman2016Giant,abdelshafy2018electron,figotin2020analytic,rouhi2021exceptional,abdelshafy2021multitransmission,figotin2021exceptional,figotin2022analytic,figotin2023analytic}.
Considering the modal propagation in the equivalent TL, the $z$-component
of the a.c. electric field induced on the cold SWS was related to
the phenomenological coupling strength coefficient $a$ as \cite{tamma2014extension,rouhi2021exceptional}

\begin{equation}
E_{z}=-a\frac{dV\left(z\right)}{dz}.\label{eq:E_z}
\end{equation}
The equivalent voltage on the TL is related to the per-unit length
impedance and equivalent current as $dV\left(z\right)/dz=-ZI\left(z\right)$.
For a lossless TL, the per-unit-length impedance is calculated by
$Z=j\beta_{\mathrm{c}}Z_{\mathrm{c}}$. Then, we relate the equivalent
voltage and current of the TL via the characteristic impedance by

\begin{equation}
\frac{dV\left(z\right)}{dz}=-j\beta_{\mathrm{c}}Z_{\mathrm{c}}I\left(z\right).\label{eq:dvdz}
\end{equation}
Substituting (\ref{eq:dvdz}) in (\ref{eq:E_z}), we obtain the relation
between the axial electric field of the guided mode and the equivalent
current of TL by

\begin{equation}
E_{z}=ja\beta_{\mathrm{c}}Z_{\mathrm{c}}I\left(z\right).\label{eq:E_z_2}
\end{equation}

Then, the interaction impedance $Z_{\mathrm{P}}$ is calculated by
(\ref{eq:Zpierce}) for the interacting harmonic (i.e., $m=1$). Here,
we derive the coupling strength coefficient in terms of $Z_{\mathrm{c}}$
and $Z_{\mathrm{P}}$. By substituting $E_{z}$ from (\ref{eq:E_z_2})
and time-average power along the TL $P=Z_{\mathrm{c}}\left|I\left(z\right)\right|^{2}/2$
in (\ref{eq:Zpierce}), the interaction impedance and characteristic
impedance of the SWS are related through the coupling strength coefficient
$a$, as $a=\sqrt{Z_{\mathrm{P}}/Z_{\mathrm{c}}}$. Using this relation
between the characteristic impedance and the interaction impedance,
one can transform the TL equivalent voltage and current of the state
vector and system matrix in (\ref{eq: System Matrix}) to be in terms
of scaled state vector quantities $V^{\prime}\left(z\right)=aV\left(z\right)$
and $I^{\prime}\left(z\right)=I\left(z\right)/a$ that maintain the
average power definition $P=\frac{1}{2}\iint_{S}\mathrm{Re}\left(-E_{y}H_{x,}^{*}\right)dxdy=\mathrm{Re}\left[VI^{*}\right]/2=\mathrm{Re}\left[V^{\prime}I^{\prime}{}^{*}\right]/2$,
where $^{*}$ is the complex conjugate operator. By making this transformation,
the system equations are expressed as

\begin{equation}
\frac{d}{dz}\boldsymbol{\Psi}^{\prime}\left(z\right)=j\mathbf{\underline{M}}^{\prime}\boldsymbol{\Psi}^{\prime}\left(z\right),
\end{equation}

\noindent where the transformed state vector is defined as $\boldsymbol{\Psi}^{\prime}\left(z\right)=\left[V^{\prime},I^{\prime},V_{\mathrm{b}},I_{\mathrm{b}}\right]^{\mathrm{T}}$
and the transformed system matrix is expressed in terms of interaction
impedance rather than characteristic impedance as

\begin{equation}
\mathbf{\underline{M}}^{\prime}=\left[\begin{array}{rrrr}
0 & a^{2}\beta_{\mathrm{c}}Z_{\mathrm{c}} & 0 & 0\\
\beta_{\mathrm{c}}/\left(a^{2}Z_{\mathrm{c}}\right) & 0 & -g & -\beta_{0}\\
0 & a^{2}\beta_{\mathrm{c}}Z_{\mathrm{c}} & \beta_{0} & \zeta_{\mathrm{sc}}\\
0 & 0 & g & \beta_{0}
\end{array}\right],
\end{equation}
or equivalently

\begin{equation}
\mathbf{\underline{M}}^{\prime}=\left[\begin{array}{rrrr}
0 & \beta_{\mathrm{c}}Z_{\mathrm{P}} & 0 & 0\\
\beta_{\mathrm{c}}/Z_{\mathrm{P}} & 0 & -g & -\beta_{0}\\
0 & \beta_{\mathrm{c}}Z_{\mathrm{P}} & \beta_{0} & \zeta_{\mathrm{sc}}\\
0 & 0 & g & \beta_{0}
\end{array}\right],\label{eq: System Matrix Trans}
\end{equation}
where the coupling strength coefficient \textit{$a$} is not present
explicitly anymore. This alternative formulation for the TWT matrix
is very informative, since the interaction impedance can be readily
found for a realistic serpentine waveguide SWS using full-wave eigenmode
simulations, i.e., by performing a simulation of only one unit cell
of the cold SWS. Furthermore, to improve the accuracy of our calculations,
we consider the ``effective interaction impedance $Z_{\mathrm{P,e}}$''
discussed in Section \ref{sec:Pierce-Impedance} by adding the correction
factor $\delta_{\mathrm{e}}$ that accounts for the nonuniform cross
sectional distribution of the electric field in the interaction area
(see Appendix \ref{app:Non-uniform-Interaction}), given by

\begin{equation}
Z_{\mathrm{P,e}}=\left(1+\delta_{\mathrm{e}}\right)^{2}Z_{\mathrm{P}}.
\end{equation}
Accordingly, the definition of the coupling strength coefficient becomes
$a=\sqrt{Z_{\mathrm{P,e}}/Z_{\mathrm{c}}}$, also reported in (\ref{eq:a-MainDef}).
Consequently, the transformed system matrix (\ref{eq: System Matrix Trans})
is finally rewritten as

\begin{equation}
\mathbf{\underline{M}}^{\prime}=\left[\begin{array}{rrrr}
0 & \beta_{\mathrm{c}}Z_{\mathrm{P,e}} & 0 & 0\\
\beta_{\mathrm{c}}/Z_{\mathrm{P,e}} & 0 & -g & -\beta_{0}\\
0 & \beta_{\mathrm{c}}Z_{\mathrm{P,e}} & \beta_{0} & \zeta_{\mathrm{sc}}\\
0 & 0 & g & \beta_{0}
\end{array}\right].
\end{equation}
The coupling strength coefficient \textit{$a$} has been eliminated
through the proposed transformation, and we can use the effective
interaction impedance $Z_{\mathrm{\mathrm{P,e}}}$ instead of the
characteristic impedance $Z_{\mathrm{c}}$ in our derived equations.
We can use this alternative definition when dealing with power since
$P=\mathrm{Re}\left[VI^{*}\right]/2=\mathrm{Re}\left[V^{\prime}I^{\prime}{}^{*}\right]/2$.
One could also use the impedance to calculate the output power as
$P_{\mathrm{out}}=\left|V^{\mathrm{o}}\right|^{2}/\left(2Z_{\mathrm{c}}\right)=\left|V^{\mathrm{o}\prime}\right|^{2}/\left(2Z_{\mathrm{P,e}}\right)$,
where $Z_{\mathrm{P,e}}=a^{2}Z_{\mathrm{c}}$ and $V^{\mathrm{o}\prime}=aV^{o}$
, assuming the TWT is matched to the modal characteristic impedance
$Z_{\mathrm{c}}$ (see Subsection \ref{subsec:Gain-Versus-Frequency}).

\section{Plasma Frequency Reduction Factor\label{app:Plasma-Reduction-Factor}}

As explained in \cite{ramo1939electronic,branch1955plasma,datta2009simple},
the finite cross section of the e-beam, along with the surrounding
metallic walls of the tunnel will make the scalar electric potential
of the e-beam nonuniform over the beam cross section. Consequently,
the plasma frequency of the beam will be reduced by the plasma frequency
reduction factor. The closed-form frequency-dependent value we use
for $R_{\mathrm{sc}}$ is calculated as \cite{datta2009simple}

\noindent 
\begin{equation}
R_{\mathrm{sc}}^{2}=1-2\mathrm{\mathit{I}_{1}}\left(\beta_{0}r_{\mathrm{b}}\right)\left(\mathrm{\mathit{K}_{1}}\left(\beta_{0}r_{\mathrm{b}}\right)+\frac{\mathrm{\mathit{K}_{0}}\left(\beta_{0}r_{\mathrm{c}}\right)}{\mathrm{\mathit{I}_{0}}\left(\beta_{0}r_{\mathrm{c}}\right)}\mathrm{\mathit{I}_{1}}\left(\beta_{0}r_{\mathrm{b}}\right)\right),\label{eq:Rsc}
\end{equation}

\noindent where, we assume the beam has a cylindrical cross section
with radius $r_{\mathrm{b}}$ and the beam tunnel is assumed to be
a metallic cylinder with a radius of $r_{\mathrm{c}}$. In addition,
$I_{n}$ and $K_{n}$ are modified Bessel functions of the first and
second kind, respectively. Moreover, the analytical method for calculating
the reduced plasma frequency based on 3D PIC simulations is developed
in \cite{mealy2022reduced} which can be used for cylindrical-shaped
e-beam flowing inside of a cylindrical tunnel.

\section{Computational Burden and Simulation Time\label{sec:Computational-Burden-and}}

For PIC simulations, we used a Dell Server PER740XD with 2 processors
of Intel(R) Xeon(R) Gold 6244 central processing unit (CPU) (24.75M
Cache, 3.60 GHz) and installed 96 GB of RAM. Furthermore, the system
is equipped with a powerful graphics card, the NVIDIA Tesla V100 Volta
graphics processing unit (GPU) accelerator (with a RAM size of 32GB).
In order to provide the PIC gain results, for the example provided
in Fig. \ref{fig:HotforHotExample}(c), the total number of mesh cells
in the simulation is around 2.6 million and a steady state output
signal is obtained after a transient time of $10\:\mathrm{ns}$ elapses,
and we swept the input RF frequency from $23\:\mathrm{GHz}$ and $34\:\mathrm{GHz}$
with frequency steps of $0.1\:\mathrm{GHz}$. It took around 21 hours
to obtain the PIC gain results over the desired frequency range using
such powerful GPU acceleration in CST Studio Suite. In contrast, once
the required primary data for our model (such as the interaction impedance
and correction factor) is obtained with full-wave simulation of just
a unit cell of the cold SWS (which are not very computationally demanding),
the theoretical output gain using our model is calculated in a few
seconds. This is done by using the implemented code of our developed
model in Mathwork Matlab R2023a.

\section*{Acknowledgment}

This material is based upon work supported by the Air Force Office
of Scientific Research (AFOSR) Multidisciplinary Research Program
of the University Research Initiative (MURI) under Grant No. FA9550-20-1-0409
administered through the University of New Mexico. The authors are
thankful to DS SIMULIA for providing CST Studio Suite that was instrumental
in this study.



\begin{thebibliography}{10}
	\providecommand{\url}[1]{#1}
	\csname url@samestyle\endcsname
	\providecommand{\newblock}{\relax}
	\providecommand{\bibinfo}[2]{#2}
	\providecommand{\BIBentrySTDinterwordspacing}{\spaceskip=0pt\relax}
	\providecommand{\BIBentryALTinterwordstretchfactor}{4}
	\providecommand{\BIBentryALTinterwordspacing}{\spaceskip=\fontdimen2\font plus
		\BIBentryALTinterwordstretchfactor\fontdimen3\font minus
		\fontdimen4\font\relax}
	\providecommand{\BIBforeignlanguage}[2]{{%
			\expandafter\ifx\csname l@#1\endcsname\relax
			\typeout{** WARNING: IEEEtran.bst: No hyphenation pattern has been}%
			\typeout{** loaded for the language `#1'. Using the pattern for}%
			\typeout{** the default language instead.}%
			\else
			\language=\csname l@#1\endcsname
			\fi
			#2}}
	\providecommand{\BIBdecl}{\relax}
	\BIBdecl
	
	\bibitem{han2005investigations}
	S.-T. Han, J.-K. So, K.-H. Jang, Y.-M. Shin, J.-H. Kim, S.-S. Chang, N.~M.
	Ryskin, and G.-S. Park, ``Investigations on a microfabricated {FWTWT}
	oscillator,'' \emph{IEEE transactions on electron devices}, vol.~52, no.~5,
	pp. 702--708, 2005.
	
	\bibitem{wong2020recent}
	P.~Wong, P.~Zhang, and J.~Luginsland, ``Recent theory of traveling-wave tubes:
	A tutorial-review,'' \emph{Plasma research express}, vol.~2, no.~2, p.
	023001, 2020.
	
	\bibitem{benford2015high}
	J.~Benford, J.~A. Swegle, and E.~Schamiloglu, \emph{High power
		microwaves}.\hskip 1em plus 0.5em minus 0.4em\relax CRC press, 2015.
	
	\bibitem{paoloni2021millimeter}
	C.~Paoloni, D.~Gamzina, R.~Letizia, Y.~Zheng, and N.~C. Luhmann~Jr,
	``Millimeter wave traveling wave tubes for the 21st century,'' \emph{Journal
		of electromagnetic waves and applications}, vol.~35, no.~5, pp. 567--603,
	2021.
	
	\bibitem{dohler1987serpentine}
	G.~Dohler, D.~Gagne, D.~Gallagher, and R.~Moats, ``Serpentine waveguide
	{TWT},'' in \emph{1987 International electron devices meeting}.\hskip 1em
	plus 0.5em minus 0.4em\relax IEEE, 1987, pp. 485--488.
	
	\bibitem{liu1995folded}
	S.~Liu, ``Folded waveguide circuit for broadband {MM} wave {TWTs},''
	\emph{International journal of infrared and millimeter waves}, vol.~16, pp.
	809--815, 1995.
	
	\bibitem{collins1998technique}
	C.~Collins, R.~Miles, R.~Pollard, D.~Steenson, J.~Digby, G.~Parkhurst,
	J.~Chamberlain, N.~Cronin, S.~Davies, and J.~W. Bowen, ``Technique for
	micro-machining millimetre-wave rectangular waveguide,'' \emph{Electronics
		Letters}, vol.~34, no.~10, pp. 996--997, 1998.
	
	\bibitem{collins1999new}
	C.~Collins, R.~Miles, J.~Digby, G.~Parkhurst, R.~Pollard, J.~Chamberlain,
	D.~Steenson, N.~Cronin, S.~Davies, and J.~W. Bowen, ``A new micro-machined
	millimeter-wave and terahertz snap-together rectangular waveguide
	technology,'' \emph{IEEE microwave and guided wave letters}, vol.~9, no.~2,
	pp. 63--65, 1999.
	
	\bibitem{gong2011experimental}
	H.~Gong, Y.~Gong, T.~Tang, J.~Xu, and W.-X. Wang, ``Experimental investigation
	of a high-power ka-band folded waveguide traveling-wave tube,'' \emph{IEEE
		transactions on electron devices}, vol.~58, no.~7, pp. 2159--2163, 2011.
	
	\bibitem{he2010investigation}
	J.~He, Y.~Wei, Z.~Lu, Y.~Gong, and W.~Wang, ``Investigation of a ridge-loaded
	folded-waveguide slow-wave system for the millimeter-wave traveling-wave
	tube,'' \emph{IEEE transactions on plasma science}, vol.~38, no.~7, pp.
	1556--1562, 2010.
	
	\bibitem{liao2010rectangular}
	M.~Liao, Y.~Wei, Y.~Gong, J.~He, W.~Wang, and G.-S. Park, ``A rectangular
	groove-loaded folded waveguide for millimeter-wave traveling-wave tubes,''
	\emph{IEEE transactions on plasma sience}, vol.~38, no.~7, pp. 1574--1578,
	2010.
	
	\bibitem{tian2011novel}
	Y.~Tian, L.~Yue, J.~Xu, W.~Wang, Y.~Wei, Y.~Gong, and J.~Feng, ``A novel
	slow-wave structure-folded rectangular groove waveguide for millimeter-wave
	{TWT},'' \emph{IEEE transactions on electron devices}, vol.~59, no.~2, pp.
	510--515, 2011.
	
	\bibitem{hou2013novel}
	Y.~Hou, Y.~Gong, J.~Xu, S.~Wang, Y.~Wei, L.~Yue, and J.~Feng, ``A novel
	ridge-vane-loaded folded-waveguide slow-wave structure for 0.22-{THz}
	traveling-wave tube,'' \emph{IEEE transactions on electron devices}, vol.~60,
	no.~3, pp. 1228--1235, 2013.
	
	\bibitem{wei2014novel}
	Y.~Wei, G.~Guo, Y.~Gong, L.~Yue, G.~Zhao, L.~Zhang, C.~Ding, T.~Tang, M.~Huang,
	W.~Wang \emph{et~al.}, ``Novel {W}-band ridge-loaded folded waveguide
	traveling wave tube,'' \emph{IEEE electron device letters}, vol.~35, no.~10,
	pp. 1058--1060, 2014.
	
	\bibitem{marosi2022Three}
	R.~Marosi, T.~Mealy, A.~Figotin, and F.~Capolino, ``Three-way serpentine slow
	wave structures with stationary inflection point and enhanced interaction
	impedance,'' \emph{IEEE Transactions on Plasma Science}, vol.~50, no.~12, pp.
	4820--4833, 2022.
	
	\bibitem{choi1995folded}
	J.~Choi, C.~Armstrong, A.~Ganguly, and F.~Calise, ``Folded waveguide gyrotron
	traveling-wave-tube amplifier,'' \emph{Physics of Plasmas}, vol.~2, no.~3,
	pp. 915--922, 1995.
	
	\bibitem{ha1998theoretical}
	H.-J. Ha, S.-S. Jung, and G.-S. Park, ``Theoretical study for folded waveguide
	traveling wave tube,'' \emph{International journal of infrared and millimeter
		waves}, vol.~19, no.~9, pp. 1229--1245, 1998.
	
	\bibitem{booske2005accurate}
	J.~H. Booske, M.~C. Converse, C.~L. Kory, C.~T. Chevalier, D.~A. Gallagher,
	K.~E. Kreischer, V.~O. Heinen, and S.~Bhattacharjee, ``Accurate parametric
	modeling of folded waveguide circuits for millimeter-wave traveling wave
	tubes,'' \emph{IEEE transactions on electron devices}, vol.~52, no.~5, pp.
	685--694, 2005.
	
	\bibitem{antonsen2013transmission}
	T.~M. Antonsen, A.~N. Vlasov, D.~P. Chernin, I.~A. Chernyavskiy, and B.~Levush,
	``Transmission line model for folded waveguide circuits,'' \emph{IEEE
		transactions on electron devices}, vol.~60, no.~9, pp. 2906--2911, 2013.
	
	\bibitem{chernin20141}
	D.~Chernin, T.~M. Antonsen, A.~N. Vlasov, I.~A. Chernyavskiy, K.~T. Nguyen, and
	B.~Levush, ``1-{D} large signal model of folded-waveguide traveling wave
	tubes,'' \emph{IEEE Transactions on Electron Devices}, vol.~61, no.~6, pp.
	1699--1706, 2014.
	
	\bibitem{chernyavskiy2013parallel}
	I.~A. Chernyavskiy, A.~N. Vlasov, B.~Levush, T.~M. Antonsen, and K.~T. Nguyen,
	``Parallel 2{D} large-signal modeling of cascaded {TWT} amplifiers,'' in
	\emph{2013 IEEE 14th international vacuum electronics conference
		(IVEC)}.\hskip 1em plus 0.5em minus 0.4em\relax IEEE, 2013, pp. 1--2.
	
	\bibitem{chernyavskiy2016large}
	I.~A. Chernyavskiy, T.~M. Antonsen, A.~N. Vlasov, D.~Chernin, K.~T. Nguyen, and
	B.~Levush, ``Large-signal 2-{D} modeling of folded-waveguide traveling wave
	tubes,'' \emph{IEEE transactions on electron devices}, vol.~63, no.~6, pp.
	2531--2537, 2016.
	
	\bibitem{meyne2016large}
	S.~Meyne, P.~Bernadi, P.~Birtel, J.-F. David, and A.~F. Jacob, ``Large-signal
	2.5-{D} steady-state beam-wave interaction simulation of folded-waveguide
	traveling-wave tubes,'' \emph{IEEE transactions on electron devices},
	vol.~63, no.~12, pp. 4961--4967, 2016.
	
	\bibitem{yan20163}
	W.-Z. Yan, Y.-L. Hu, Y.-X. Tian, J.-Q. Li, and B.~Li, ``A 3-{D} large-signal
	model of folded-waveguide {TWTs},'' \emph{IEEE Transactions on Electron
		Devices}, vol.~63, no.~2, pp. 819--826, 2016.
	
	\bibitem{li2015nonlinear}
	K.~Li, W.~Liu, Y.~Wang, and M.~Cao, ``A nonlinear analysis of the terahertz
	serpentine waveguide traveling-wave amplifier,'' \emph{Physics of plasmas},
	vol.~22, no.~4, p. 043115, 2015.
	
	\bibitem{zhang2020active}
	R.~Zhang, X.~Lin, T.~Wang, X.~Xiao, Z.~Wang, Z.~Duan, Y.~Gong, J.~Feng,
	G.~Travish, and H.~Gong, ``An active transmission matrix-based nonlinear
	analysis for folded waveguide {TWT},'' \emph{IEEE transactions on electron
		devices}, vol.~67, no.~3, pp. 1205--1210, 2020.
	
	\bibitem{figotin2023analytic}
	A.~Figotin, ``Analytic theory of coupled-cavity traveling wave tubes,''
	\emph{Journal of mathematical physics}, vol.~64, no.~4, p. 042705, 2023.
	
	\bibitem{pierce1947theory}
	J.~R. Pierce, ``Theory of the beam-type traveling-wave tube,''
	\emph{Proceedings of the IRE}, vol.~35, no.~2, pp. 111--123, 1947.
	
	\bibitem{pierce1949new}
	J.~R. Pierce and W.~B. Hebenstreit, ``A new type of high-frequency amplifier,''
	\emph{The Bell system technical journal}, vol.~28, no.~1, pp. 33--51, 1949.
	
	\bibitem{pierce1950Book}
	J.~R. Pierce, \emph{Traveling-wave tubes}.\hskip 1em plus 0.5em minus
	0.4em\relax D.Van Nostrand Company Inc., 1950.
	
	\bibitem{pierce1951waves}
	------, ``Waves in electron streams and circuits,'' \emph{Bell System Technical
		Journal}, vol.~30, no.~3, pp. 626--651, 1951.
	
	\bibitem{rouhi2021exceptional}
	K.~Rouhi, R.~Marosi, T.~Mealy, A.~F. Abdelshafy, A.~Figotin, and F.~Capolino,
	``Exceptional degeneracies in traveling wave tubes with dispersive slow-wave
	structure including space-charge effect,'' \emph{Applied physics letters},
	vol. 118, no.~26, p. 263506, 2021.
	
	\bibitem{mealy2022traveling}
	T.~Mealy and F.~Capolino, ``Traveling wave tube eigenmode solution for
	beam-loaded slow wave structure based on particle-in-cell simulations,''
	\emph{IEEE transactions on plasma science}, vol.~50, no.~3, pp. 635--648,
	2022.
	
	\bibitem{figotin2021exceptional}
	A.~Figotin, ``Exceptional points of degeneracy in traveling wave tubes,''
	\emph{Journal of mathematical physics}, vol.~62, no.~8, p. 082701, 2021.
	
	\bibitem{marcuvitz1951waveguide}
	N.~Marcuvitz, \emph{Waveguide handbook}.\hskip 1em plus 0.5em minus 0.4em\relax
	IET, 1951.
	
	\bibitem{collin2001foundations}
	R.~E. Collin, ``Foundations for microwave engineering,'' 2001.
	
	\bibitem{bhattacharjee2004folded}
	S.~Bhattacharjee, J.~H. Booske, C.~L. Kory, D.~W. Van Der~Weide, S.~Limbach,
	S.~Gallagher, J.~D. Welter, M.~R. Lopez, R.~M. Gilgenbach, R.~L. Ives
	\emph{et~al.}, ``Folded waveguide traveling-wave tube sources for terahertz
	radiation,'' \emph{IEEE transactions on plasma science}, vol.~32, no.~3, pp.
	1002--1014, 2004.
	
	\bibitem{sumathy2013design}
	M.~Sumathy, D.~Augustin, S.~K. Datta, L.~Christie, and L.~Kumar, ``Design and
	{RF} characterization of {W}-band meander-line and folded-waveguide slow-wave
	structures for {TWTs},'' \emph{IEEE transactions on electron devices},
	vol.~60, no.~5, pp. 1769--1775, 2013.
	
	\bibitem{hung2015absolute}
	D.~Hung, I.~Rittersdorf, P.~Zhang, D.~Chernin, Y.~Lau, T.~Antonsen~Jr,
	J.~Luginsland, D.~Simon, and R.~Gilgenbach, ``Absolute instability near the
	band edge of traveling-wave amplifiers,'' \emph{Physical review letters},
	vol. 115, no.~12, p. 124801, 2015.
	
	\bibitem{felsen1994radiation}
	L.~B. Felsen and N.~Marcuvitz, \emph{Radiation and scattering of waves}.\hskip
	1em plus 0.5em minus 0.4em\relax John Wiley \& Sons, 1994.
	
	\bibitem{liu2000study}
	S.~Liu, ``Study of propagating characteristics for folded waveguide {TWT} in
	millimeter wave,'' \emph{International journal of infrared and millimeter
		waves}, vol.~21, no.~4, pp. 655--660, 2000.
	
	\bibitem{li2013dispersion}
	K.~Li, W.~Liu, Y.~Wang, and M.~Cao, ``Dispersion characteristics of two-beam
	folded waveguide for terahertz radiation,'' \emph{IEEE transactions on
		electron devices}, vol.~60, no.~12, pp. 4252--4257, 2013.
	
	\bibitem{nguyen2014design}
	K.~T. Nguyen, A.~N. Vlasov, L.~Ludeking, C.~D. Joye, A.~M. Cook, J.~P. Calame,
	J.~A. Pasour, D.~E. Pershing, E.~L. Wright, S.~J. Cooke \emph{et~al.},
	``Design methodology and experimental verification of
	serpentine/folded-waveguide {TWTs},'' \emph{IEEE transactions on electron
		devices}, vol.~61, no.~6, pp. 1679--1686, 2014.
	
	\bibitem{meyne2017simulation}
	S.~Meyne, \emph{Simulation and design of traveling-wave tubes with
		folded-waveguide delay lines}.\hskip 1em plus 0.5em minus 0.4em\relax
	Technische Universit{\"a}t Hamburg-Harburg, 2017.
	
	\bibitem{na2002analysis}
	Y.~H. Na, S.~W. Chung, and J.~J. Choi, ``Analysis of a broadband q band folded
	waveguide traveling-wave tube,'' \emph{IEEE transactions on plasma science},
	vol.~30, no.~3, pp. 1017--1023, 2002.
	
	\bibitem{han2003design}
	S.-T. Han, J.-I. Kim, and G.-S. Park, ``Design of a folded waveguide
	traveling-wave tube,'' \emph{Microwave and Optical Technology Letters},
	vol.~38, no.~2, pp. 161--165, 2003.
	
	\bibitem{carter2018microwave}
	R.~G. Carter, \emph{Microwave and RF vacuum electronic power sources}.\hskip
	1em plus 0.5em minus 0.4em\relax Cambridge University Press, 2018.
	
	\bibitem{zheng2009parametric}
	R.~Zheng and X.~Chen, ``Parametric simulation and optimization of cold-test
	properties for a 220 ghz broadband folded waveguide traveling-wave tube,''
	\emph{Journal of infrared, millimeter, and terahertz waves}, vol.~30, no.~9,
	pp. 945--958, 2009.
	
	\bibitem{gewartowski1965principles}
	J.~W. Gewartowski and H.~A. Watson, \emph{Principles of electron tubes:
		including grid-controlled tubes, microwave tubes, and gas tubes}.\hskip 1em
	plus 0.5em minus 0.4em\relax Van Nostrand, 1965.
	
	\bibitem{sharma2014design}
	R.~K. Sharma, A.~Grede, S.~Chaudhary, V.~Srivastava, and H.~Henke, ``Design of
	folded waveguide slow-wave structure for {W}-band {TWT},'' \emph{IEEE
		transactions on plasma science}, vol.~42, no.~10, pp. 3430--3436, 2014.
	
	\bibitem{sudhamani2017investigation}
	H.~Sudhamani, J.~Balakrishnan, and S.~Reddy, ``Investigation of instabilities
	in a folded-waveguide sheet-beam {TWT},'' \emph{IEEE transactions on electron
		devices}, vol.~64, no.~10, pp. 4266--4271, 2017.
	
	\bibitem{gilmour1994principles}
	A.~Gilmour, \emph{Principles of traveling wave tubes}.\hskip 1em plus 0.5em
	minus 0.4em\relax Artech House Radar Library, 1994.
	
	\bibitem{tamma2014extension}
	V.~A. Tamma and F.~Capolino, ``Extension of the pierce model to multiple
	transmission lines interacting with an electron beam,'' \emph{IEEE
		transactions on plasma science}, vol.~42, no.~4, pp. 899--910, 2014.
	
	\bibitem{rouhi2023modeling}
	K.~Rouhi, R.~Marosi, T.~Mealy, A.~Figotin, and F.~Capolino, ``Modeling of
	serpentine waveguide traveling wave tube to calculate gain diagram,'' in
	\emph{2023 24th International vacuum electronics conference (IVEC)}.\hskip
	1em plus 0.5em minus 0.4em\relax IEEE, 2023, pp. 1--2.
	
	\bibitem{figotin2013multi}
	A.~Figotin and G.~Reyes, ``Multi-transmission-line-beam interactive system,''
	\emph{Journal of mathematical physics}, vol.~54, no.~11, p. 111901, 2013.
	
	\bibitem{figotin2020analytic}
	A.~Figotin, \emph{An analytic theory of multi-stream electron beams in
		traveling wave tubes}.\hskip 1em plus 0.5em minus 0.4em\relax World
	Scientific, 2020.
	
	\bibitem{marcuvitz1951representation}
	N.~Marcuvitz and J.~Schwinger, ``On the representation of the electric and
	magnetic fields produced by currents and discontinuities in wave guides.
	{I},'' \emph{Journal of applied physics}, vol.~22, no.~6, pp. 806--819, 1951.
	
	\bibitem{hammer1967coupling}
	J.~Hammer, ``Coupling between slow waves and convective instabilities in
	solids,'' \emph{Applied physics letters}, vol.~10, no.~12, pp. 358--360,
	1967.
	
	\bibitem{branch1955plasma}
	G.~Branch and T.~Mihran, ``Plasma frequency reduction factors in electron
	beams,'' \emph{IRE transactions on electron devices}, vol.~2, no.~2, pp.
	3--11, 1955.
	
	\bibitem{booske2004insights}
	J.~H. Booske and M.~C. Converse, ``Insights from one-dimensional linearized
	pierce theory about wideband traveling-wave tubes with high space charge,''
	\emph{IEEE transactions on plasma science}, vol.~32, no.~3, pp. 1066--1072,
	2004.
	
	\bibitem{antonsen1998traveling}
	T.~Antonsen and B.~Levush, ``Traveling-wave tube devices with nonlinear
	dielectric elements,'' \emph{IEEE transactions on plasma science}, vol.~26,
	no.~3, pp. 774--786, 1998.
	
	\bibitem{he2011investigation}
	J.~He, Y.~Wei, Y.~Gong, W.~Wang, and G.-S. Park, ``Investigation on a {W} band
	ridge-loaded folded waveguide {TWT},'' \emph{IEEE transactions on plasma
		science}, vol.~39, no.~8, pp. 1660--1664, 2011.
	
	\bibitem{hou2012equivalent}
	Y.~Hou, J.~Xu, H.-R. Yin, Y.-Y. Wei, L.-N. Yue, G.~Zhao, and Y.-B. Gong,
	``Equivalent circuit analysis of ridge-loaded folded-waveguide slow-wave
	structures for millimeter-wave traveling-wave tubes,'' \emph{Progress in
		electromagnetics research}, vol. 129, pp. 215--229, 2012.
	
	\bibitem{guo2012tapered}
	G.~Guo, Y.~Wei, L.~Yue, Y.~Gong, X.~Xu, J.~He, G.~Zhao, W.~Wang, and G.-S.
	Park, ``A tapered ridge-loaded folded waveguide slow-wave structure for
	millimeter-wave traveling-wave tube,'' \emph{Journal of infrared, millimeter,
		and terahertz waves}, vol.~33, pp. 131--140, 2012.
	
	\bibitem{Othman2016Theory}
	M.~A.~K. Othman, V.~A. Tamma, and F.~Capolino, ``Theory and new amplification
	regime in periodic multimodal slow wave structures with degeneracy
	interacting with an electron beam,'' \emph{IEEE transactions on plasma
		science}, vol.~44, no.~4, pp. 594--611, Apr 2016.
	
	\bibitem{Othman2016Giant}
	M.~A.~K. Othman, M.~Veysi, A.~Figotin, and F.~Capolino, ``Giant amplification
	in degenerate band edge slow-wave structures interacting with an electron
	beam,'' \emph{Physics of plasmas}, vol.~23, no.~3, p. 033112, Mar 2016.
	
	\bibitem{abdelshafy2018electron}
	A.~F. Abdelshafy, M.~A. Othman, F.~Yazdi, M.~Veysi, A.~Figotin, and
	F.~Capolino, ``Electron-beam-driven devices with synchronous multiple
	degenerate eigenmodes,'' \emph{IEEE transactions on plasma science}, vol.~46,
	no.~8, pp. 3126--3138, 2018.
	
	\bibitem{abdelshafy2021multitransmission}
	A.~F. Abdelshafy, M.~A. Othman, A.~Figotin, and F.~Capolino,
	\emph{Multitransmission line model for slow wave structures interacting with
		electron beams and multimode synchronization}.\hskip 1em plus 0.5em minus
	0.4em\relax John Wiley \& Sons, Ltd, 2021, pp. 17--56.
	
	\bibitem{figotin2022analytic}
	A.~Figotin, ``Analytic theory of multicavity klystrons,'' \emph{Journal of
		mathematical physics}, vol.~63, no.~6, p. 062703, 2022.
	
	\bibitem{ramo1939electronic}
	S.~Ramo, ``The electronic-wave theory of velocity-modulation tubes,''
	\emph{Proceedings of the IRE}, vol.~27, no.~12, pp. 757--763, 1939.
	
	\bibitem{datta2009simple}
	S.~K. Datta and L.~Kumar, ``A simple closed-form formula for plasma-frequency
	reduction factor for a solid cylindrical electron beam,'' \emph{IEEE
		transactions on electron devices}, vol.~56, no.~6, pp. 1344--1346, 2009.
	
	\bibitem{mealy2022reduced}
	T.~Mealy, R.~Marosi, and F.~Capolino, ``Reduced plasma frequency calculation
	based on particle-in-cell simulations,'' \emph{IEEE Transactions on Plasma
		Science}, vol.~50, no.~10, pp. 3570--3577, 2022.
	
\end{thebibliography}
\end{document}